\documentclass[aps,pra,10pt,notitlepage,nofootinbib,tightenlines,floatfix,
twocolumn,superscriptaddress]{revtex4-2}

\usepackage[english]{babel}
\usepackage[margin=50pt]{geometry}

\usepackage[colorlinks=true, allcolors=blue]{hyperref}
\hypersetup{
    colorlinks=true,       
    linkcolor=red,          
    citecolor=magenta,        
    filecolor=magenta,      
    urlcolor=cyan,           
    runcolor=cyan
}
\usepackage{bm,amssymb,amsmath,amsfonts,times}
\usepackage{natbib}
\usepackage{graphicx}
\usepackage{color}
\usepackage{bbold}
\usepackage[utf8]{inputenc}
\usepackage{physics}

\newcommand{\eq}[1]{\begin{align}#1\end{align}}
\newcommand{\seq}[1]{\begin{align}\begin{split}#1\end{split}\end{align}}
\newcommand{\mrm}{\mathrm}
\newcommand{\id}{\mathbb{1}}
\renewcommand{\l}{\left}
\renewcommand{\r}{\right}
\newcommand{\Hred}{H_{p}}
\newcommand{\tstar}{T_2^*}
\newcommand{\axH}{a_{x}^{H}}
\newcommand{\ayH}{a_{y}^{H}}
\newcommand{\Bperp}{B_{\perp}}
\newcommand{\Bpar}{B_{\parallel}}
\newcommand{\ueV}{\,\mu\mathrm{eV}}
\newcommand{\nm}{\,\mathrm{nm}}

\graphicspath{ {figures/} }

\begin{document}
 
\title{Noise-Robust Spin-Orbit Qubit in Germanium Holes via p-Orbital Encoding}
 
\author{Yasuo Oda}
\affiliation{Department of Physics, University of Maryland Baltimore County (UMBC), Baltimore, Maryland 21250, USA}
 
\author{Jason P.\ Kestner}
\affiliation{Department of Physics, University of Maryland Baltimore County (UMBC), Baltimore, Maryland 21250, USA}
 
\date{\today}
 
\begin{abstract}
Germanium hole spin qubits are a leading platform for semiconductor quantum computation due to their strong spin-orbit coupling, all-electrical operability, and absence of valley degeneracy. 
A central obstacle is charge noise, which couples to the qubit through the same spin-orbit interaction that enables fast electrical control. 
In this work, we propose a new operational mode of a three-hole quantum dot in a planar Ge/SiGe heterostructure, modeled within a four-band Luttinger-Kohn--Bir-Pikus framework: a spin-$p$-orbital (SpO) qubit encoded in the $p$-shell of the topmost hole. 
We characterize charge noise sweet spots in the parameter space of electrostatic confinement and magnetic field, and estimate that relaxation rates of the SpO qubit are comparable to those of spin qubits hosted in single holes. 
We then design and optimize an all-electrical Landau-Zener state-transfer protocol that induces logical qubit state transitions without microwave driving, and we show that the quadrupole-quadrupole Coulomb interaction between neighboring dots enables fast two-qubit entangling gates operated by adiabatic shuttling.
\end{abstract}
 
\maketitle

\section{Introduction}
\label{sec:intro}

Spins confined in semiconductor quantum dots are a leading candidate for scalable quantum computation~\cite{Burkard2023,Xue2022}. Germanium, in particular, has emerged as a compelling host material~\cite{Scappucci2020}. On the one hand, the strong spin-orbit coupling (SOC) of valence band holes enables all-electrical qubit manipulation via electric-dipole spin resonance (EDSR) without micromagnets~\cite{Watzinger2018,Hendrickx2020}. In addition, gate operations approaching the fault-tolerant threshold have been demonstrated~\cite{Lawrie2023,Hendrickx2021b}, and the absence of valley degeneracy removes a key source of leakage present in silicon~\cite{Scappucci2020,Terrazos2021}.

Despite this rapid progress, charge noise remains the central bottleneck for scaling planar Ge hole qubits~\cite{Lodari2021,Hendrickx2024}. 
The same SOC that enables fast gates also couples the qubit frequency to electric-field fluctuations, limiting dephasing times to $\tstar \sim 100~\mathrm{ns}$--$10~\mu\mathrm{s}$ depending on operating conditions~\cite{Hendrickx2020,Hendrickx2021b,Hendrickx2024}. 
A common strategy is to operate at so-called sweet spots, where the qubit frequency is first-order insensitive to gate-voltage fluctuations~\cite{Bosco2021,Michal2021,Wang2024,Sarkar2023,Hendrickx2024}. 
Recent theoretical efforts have focused on modeling the effect of charge noise on the spin-qubit subspace as a fluctuating gate voltage~\cite{Wang2024} or as an ensemble of two-level fluctuators (TLFs)~\cite{Wang2024,Wang2025,Kepa2023}.
These analyses build on multiband $k\cdot p$ descriptions of the strongly spin-orbit-coupled Ge valence band, which provide the analytical tools for quantitative analysis of $g$-tensors, sweet spots, and relaxation in planar Ge dots~\cite{Terrazos2021,Michal2021,Sarkar2023}.

A complementary route to noise protection is to engineer the structure of the spatial charge distribution of the qubit, and thus its coupling to the electrostatic environment.
The elementary semiconductor charge qubit, a single electron delocalized over a double dot~\cite{Gorman2005,Petersson2010}, couples to electric fields through its large dipole moment, enabling fast gates but suffering fast dephasing that persists even at its charge sweet spot~\cite{Kim2015,Stano2022}.
The charge quadrupole qubit eliminates the dipole moment altogether by encoding in symmetric triple-dot states~\cite{Friesen2017,Kornich2018,Koski2020,Kratochwil2021}, at the cost of requiring more quantum dots and low-lying leakage states. 

The benefits of higher-multipole encodings in multi-dot arrays were recognized early on~\cite{Oi2005,Hentschel2007}.
Recently, Ref.~\cite{Caporaletti2025} proposed realizing the quadrupolar coupling within a single dot: a five-electron qubit in Si, encoded in the single-particle $p$-orbital-like states, combining a single quantum dot requirement with baseband all-electrical control, named the pO qubit.
Translating this idea to planar Ge holes is natural: the valence band supplies strong SOC for electrical control, the light in-plane mass allows large dots that relax fabrication constraints~\cite{Terrazos2021}, and the mature Ge/SiGe platform offers immediate experimental access. 
However, the strong mixing between heavy- and light-holes, the anisotropic $g$-tensor, and the orbital effects of in-plane magnetic fields make the physics qualitatively different from the Si conduction band and call for a dedicated treatment~\cite{Terrazos2021,Sarkar2023,Wang2024}.

In this work, we propose hosting the logical states on the excited orbital ($p$-orbital) manifold of Ge holes, rather than the lowest spin states, and show that the additional degrees of freedom can be exploited for noise-resilient qubit encoding and all-electrical control~\cite{Caporaletti2025}. 
We model the hole physics within a shell-filling approximation, where we assume two holes fill the lowest two $s$-shell orbitals, which limits orbital relaxation channels, and the third hole occupies a $p$-orbital state. 
This shell-filling treatment neglects hole-hole interactions, which may at first seem unwarranted as they are in fact larger than the confinement energy. The true low-lying wavefunctions may even be Wigner molecule states with markedly different spatial extent and radial features than the noninteracting $p$-orbital states we assume~\cite{PerezFadon2025}. However, we expect this approximation to be qualitatively valid since, given the rotational symmetry of the Coulomb interaction, the crucial angular momentum structure (e.g., a suppressed dipole moment) is unaffected by interactions. A quantitative account of the exact interaction effects will require a full configuration-interaction treatment of the three-hole problem, which we leave for future work.

\begin{figure*}[t]
    \centering
    \includegraphics[trim={.3cm 6cm 0.5cm 0.cm}, clip,width=\textwidth]{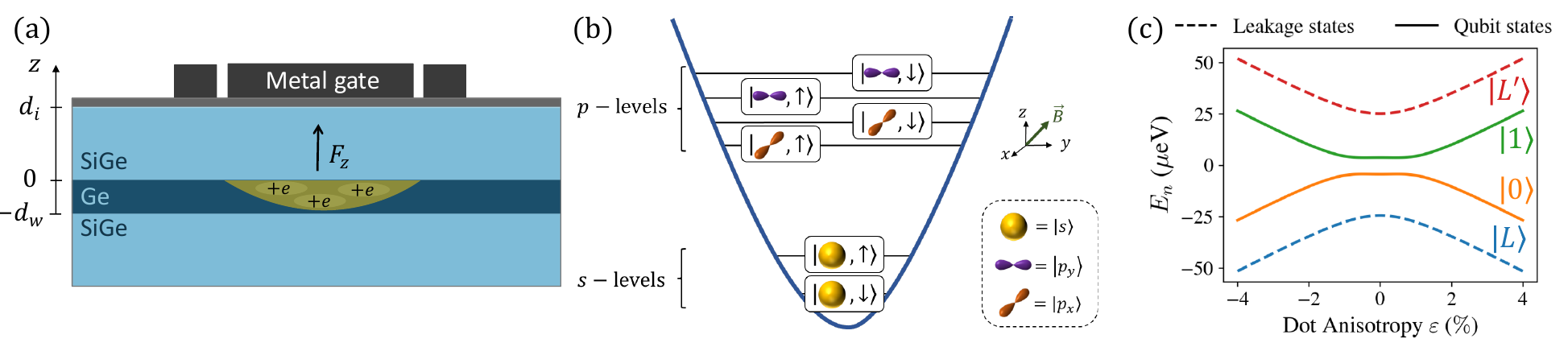}
\caption{Three-hole quantum dot in a planar Ge/SiGe heterostructure and energy level structure.
(a) Cross-sectional schematic of the gate-defined quantum dot. 
The hole is confined vertically by the SiGe/Ge/SiGe layer structure, with a Ge quantum well of width $d_w = 18~$nm, a SiGe cap layer of thickness $d_i=60~$nm, and an insulating oxide layer. 
Lateral confinement is provided electrostatically by the metal gates. 
A uniform vertical electric field $F_z$ pushes the hole toward the upper SiGe/Ge interface, and the accumulated hole wavefunction is illustrated in the quantum well.
(b) Schematic energy level diagram in the three-hole shell-filling picture. 
The $s$-shell doublet is fully occupied by two holes. 
The qubit is encoded in the single hole occupying the $p$-shell, which splits into four levels predominantly consisting of mixtures of $\{|p_x,\!+3/2\rangle,\,|p_x,\!-3/2\rangle,\,|p_y,\!+3/2\rangle,\,|p_y,\!-3/2\rangle\}$.
(c) Energy spectrum of the four $p$-shell levels as a function of the dot anisotropy $\varepsilon$. 
Solid lines denote the two logical qubit states $\ket{0}$ and $\ket{1}$, and dashed lines the two leakage states $\ket{L}$ and $\ket{L'}$. 
The nearly flat spectrum near the $\varepsilon = 0$ avoided crossing provides the qubit with robustness against noise.}
\label{fig:main}
\end{figure*}

To study the system dynamics, we use a four-band Luttinger-Kohn--Bir-Pikus (LKBP) framework applied to a realistic Ge/SiGe heterostructure~\cite{Luttinger1955,BirPikus1974}. 
We present the three-hole spin-orbit qubit proposal that encodes a logical qubit in the $p$-shell manifold at a noise-protected operating point (see Fig.~\ref{fig:main}), which we refer to as \textit{SpO qubit}.

We perform a systematic characterization of its charge-noise sweet spots, and show that coherence times are not in principle limited by phonon-induced relaxation rates. 
We propose an all-electrical Landau-Zener (LZ) state-transfer protocol that rotates the logical qubit using baseband confinement-parameter sweeps alone, optimized to eliminate leakage to spectator states. 
Lastly, we propose a two-qubit entangling gate mediated by the quadrupole-quadrupole Coulomb interaction between neighboring dots, and operated by adiabatic shuttling the dots together into an interaction zone, then separating them again. 
The shuttle rate is optimized via Makhlin local invariants to produce a controlled-$Z$ gate.
This analysis shows that high quality factors can in principle be achieved, and combined with the relatively low hardware overhead, suggests that the SpO qubit in Ge can be a promising candidate for a quantum information processing architecture.

\section{Model}
\label{sec:model}

\subsection{Hamiltonian}
\label{ssec:hamiltonian}

We consider a quantum dot in a strained Ge/SiGe planar heterostructure occupied by three holes, as depicted schematically in Fig.~\ref{fig:main}(a). 
The qubit is encoded in the $p$-shell orbital degree of freedom of the topmost hole, while the lowest two holes fill the $s$-shell and are treated as an inert, fully occupied core (shell-filling approximation)~\cite{Leon2020,Caporaletti2025}, as sketched in Fig.~\ref{fig:main}(b). 
Within this picture the system reduces to an effective single-hole problem, governed by the Hamiltonian
\seq{
  H = H_\mathrm{LK} + H_\mathrm{BP} + V_\perp(z) + V_\parallel(x,y) + H_\mathrm{Z}.
  \label{eq:Htot}
}
The kinetic term $H_\mathrm{LK}$ is described using the four-band Luttinger-Kohn (LK) Hamiltonian~\cite{Luttinger1955,Winkler2003}, corresponding to the four states with total angular momentum $j=3/2$. 
In germanium, the split-off valence band with $j=1/2$ is separated from the $j=3/2$ manifold by a large energy gap $\Delta_\mathrm{SO} \approx 0.296\,\mathrm{eV}$~\cite{Winkler2003}, justifying the truncation of the full six-band model to four bands. 
In the basis of angular momentum eigenstates $\l\{|\tfrac{3}{2},\tfrac{3}{2}\rangle,\,|\tfrac{3}{2},-\tfrac{3}{2}\rangle,\,|\tfrac{3}{2},\tfrac{1}{2}\rangle,\,|\tfrac{3}{2},-\tfrac{1}{2}\rangle\r\}$, where $\ket{j,m_j}$ denote the total angular momentum $j$ and its projection $m_j$, the LK Hamiltonian~\cite{Luttinger1955} takes the following block diagonal form,
\seq{
H_\mathrm{LK} = \begin{pmatrix}
P+Q & 0 & S & R \\
0 & P+Q & R^\dagger & -S^\dagger\\
S^\dagger & R & P-Q & 0 \\
R^\dagger & -S & 0 & P-Q
\end{pmatrix},
\label{eq:HLK}
}
where the blocks $P,Q,R,S$ depend exclusively on momentum, as specified below.
The diagonal blocks $P\pm Q$ describe the kinetic energy of the heavy-hole (HH) and light-hole (LH) states, with angular momentum projection $m_j=\pm\tfrac{3}{2}$ and $m_j=\pm\tfrac{1}{2}$ respectively, while $R$ and $S$ capture the coupling between HHs and LHs. 
The explicit forms of the block terms are
\seq{
P &= \frac{\hbar^2}{2 m_0} \gamma_1 \l(k_x^2 + k_y^2 + k_z^2 \r), \\
Q &= \frac{\hbar^2}{2 m_0} \gamma_2 \l(k_x^2 + k_y^2 - 2k_z^2 \r), \\
R &= \sqrt{3} \frac{\hbar^2}{2 m_0} \l[ - \gamma_2 \l( k_x^2-k_y^2\r)
   + i\gamma_3 \l\{k_x, k_y\r\} \r], \\
S &= -\sqrt{3} \frac{\hbar^2}{2 m_0} \gamma_3 \l\{ k_x-i k_y, k_z \r\},
\label{eq:PQRS}
}
where $k_\ell = -i \partial_\ell$ is the momentum operator along direction $\ell=x,y,z$, $\hbar$ is the reduced Planck constant, $m_0$ is the bare electron mass, and $\l\{\cdot,\cdot\r\}$ denotes the anticommutator. 
We use the Luttinger parameters $\gamma_1=13.38$, $\gamma_2=4.24$, and $\gamma_3=5.69$ for Ge~\cite{Wang2024,Terrazos2021,Lawaetz1971}, which define the pure HH/LH in-plane effective masses $m_\parallel^{H/L}=m_0/(\gamma_1\pm\gamma_2)$.

The effect of the compressive biaxial strain in the Ge quantum well is captured by the Bir-Pikus Hamiltonian $H_\mrm{BP}$~\cite{BirPikus1974}. The full Bir-Pikus Hamiltonian including shear strain, needed for the phonon coupling of Sec.~\ref{ssec:relaxation}, is given in Appendix~\ref{app:phonon}.
For static and uniform heterostructure strain, we neglect all shear components, i.e., $\epsilon_{xy}=\epsilon_{xz}=\epsilon_{yz}=0$~\cite{Sammak2019,Scappucci2020}. 
Under this approximation, the Bir-Pikus Hamiltonian becomes diagonal in the HH/LH basis,
\seq{
H_\mathrm{BP} = \mathrm{diag}\l(
  P_\epsilon+Q_\epsilon,\;P_\epsilon+Q_\epsilon,\;
  P_\epsilon-Q_\epsilon,\;P_\epsilon-Q_\epsilon
\r),
\label{eq:HBP}
}
with
\seq{
P_\epsilon &= - a_v (\epsilon_{xx}+\epsilon_{yy}+\epsilon_{zz}), \\
Q_\epsilon &= - \frac{b_v}{2} (\epsilon_{xx}+\epsilon_{yy}-2\epsilon_{zz}).
\label{eq:PQeps}
}
The in-plane and out-of-plane strain components are related by Poisson's ratio, $\epsilon_{zz} = -2(C_{12}/C_{11})\epsilon_{xx}$, with $\epsilon_{xx}=\epsilon_{yy}=-0.6\%$ and $\epsilon_{zz}=0.42\%$ for a Si$_{0.2}$Ge$_{0.8}$ barrier~\cite{Terrazos2021,Sammak2019}. 
The deformation potential constants are $a_v = 2.0\,\mathrm{eV}$ and $b_v = -2.16\,\mathrm{eV}$~\cite{VandeWalle1989}. 
It is worth noting that strain contributes directly to breaking the HH/LH degeneracy, as well as to the strength of SOC~\cite{Terrazos2021}.

The vertical confinement potential $V_\perp$ combines the offset due to the quantum-well heterostructure with the applied gate vertical electric field $F_z > 0$, used to accumulate the hole near the upper SiGe/Ge interface [Fig.~\ref{fig:main}(a)].
This serves to tune the degree of HH/LH mixing as well as the effective $g$-factor~\cite{Wang2024,Sarkar2023,Hendrickx2024}. 
The resulting potential is
\seq{
V_\perp^{\alpha}(z) = - e F_z z - \Delta U_\alpha\,\Theta(-z)\,\Theta(z + d_w),
\label{eq:Vperp}
}
where $\alpha = H, L$ denotes the HH or LH band, $\Delta U_\alpha$ is the band offset in the strained Ge quantum well, $e>0$ is the elementary charge, and $\Theta(z)$ is the Heaviside step function. 
We use $F_z=1.5~$MV/m, $\Delta U_H = 150\,\mathrm{meV}$ and $\Delta U_L = 100\,\mathrm{meV}$~\cite{Wang2024,Schffler1997}, with the quantum well of width $d_w = 18~$nm, and a SiGe region of height $d_i=60~$nm bounded above by a SiO$_2$ interface modeled as an infinite potential barrier at $z = d_i$.

In addition, an in-plane parabolic electrostatic potential provides lateral confinement,
\seq{
V_\parallel^{\alpha}(x,y) = \frac{\hbar^2}{2 m_\parallel^\alpha}
  \l[\frac{x^2}{(a_x^{\alpha})^4} + \frac{y^2}{(a_y^{\alpha})^4} \r],
\label{eq:Vpar}
}
where $a_\ell^{\alpha}$ denotes the harmonic confinement length along direction $\ell = x, y$ for band $\alpha$. 
Because the electrostatic potential must be the same for both bands, the HH and LH confinement lengths are not independent:  $m_\parallel^H \l(a_\ell^{H}\r)^4 = m_\parallel^L \l(a_\ell^{L}\r)^4$. 
The independent control parameters are therefore the HH lengths $\axH$ and $\ayH$, which set the in-plane anisotropy and average size of the dot.
We define the anisotropy of the dot as the parameter $\varepsilon$ such that $a_x^H=a_0(1+\varepsilon)$ and $a_y^H=a_0(1-\varepsilon)$, where $a_0=\l(a_x^H+a_y^H\r)/2$ is the average radius of the HH.

Finally, the Zeeman term quantifies the interaction of the hole with an external magnetic field $\vec{B} = (B_x, B_y, B_z)$,
\seq{
H_\mathrm{Z} = 2\mu_B\l( \mathcal{J}_x B_x + \mathcal{J}_y B_y + \mathcal{J}_z B_z \r),
\label{eq:Zeeman}
}
where $\mathcal{J}_\ell = \kappa J_\ell + q J_\ell^3$ with $\ell = x,y,z$, $J_\ell$ are the spin-$\tfrac{3}{2}$ matrices (given explicitly in Appendix~\ref{app:Jmatrices}), and the isotropic and anisotropic Zeeman coefficients for Ge are $\kappa = 3.41$ and $q = 0.067$~\cite{Lawaetz1971,Winkler2003}, and $\mu_B=e \hbar/2m_0$ is the Bohr magneton. 
Throughout this work, the in-plane field component is oriented at $45^\circ$ between the $x$- and $y$-axes, so we parametrize $\vec{B} = \l(\Bpar/\sqrt{2}, \Bpar/\sqrt{2}, \Bperp\r)$ and specify the field by its in-plane magnitude $\Bpar$ and out-of-plane component $\Bperp$.
We discuss other in-plane magnetic field directions in Appendix~\ref{app-sec:in-plane-B-dependence}.

\subsection{Orbital basis}
\label{ssec:orbital_basis}

The effective single-hole Hamiltonian $H$ is diagonalized by projecting onto a basis of product states that separate the in-plane and out-of-plane degrees of freedom. 
For each band index $\alpha \in \{H, L\}$, the basis states take the form
\seq{
    \Psi^{\alpha}_{n_x,n_y,k}(x,y,z)
    = \phi^{\alpha}_{n_x,n_y}(x,y)\,\psi^{\alpha}_{k}(z),
    \label{eq:basis_product}
}
where $\phi^{\alpha}_{n_x,n_y}(x,y)$ are the in-plane harmonic oscillator orbitals and $\psi^{\alpha}_{k}(z)$ are the out-of-plane sub-band wavefunctions, each described below.

The in-plane motion is governed by the lateral confinement shown in Eq.~\eqref{eq:Vpar}, rewritten as $V_\parallel^{\alpha} = \frac{1}{2}m^\alpha_\parallel \l[(\omega_x^{\alpha})^2 x^2 + (\omega_y^{\alpha})^2 y^2\r]$ with $\omega_\ell^{\alpha} = \hbar/[m^\alpha_\parallel (a_\ell^{\alpha})^2]$, and by the out-of-plane magnetic field $B_\perp$. 
In general, the effect of the magnetic field on orbital dynamics is accounted for by including the vector potential $\vec{A}$ via the minimal coupling $k_\ell\rightarrow k_\ell + (e/\hbar) A_\ell$. 
Here, we work in the Landau gauge, where $\vec{A} = \l(2 z B_y - y B_\perp, - 2 z B_x + x B_\perp, 0\r)/2$~\cite{Stano2019}.

We project the Hamiltonian onto the basis of products of one-dimensional harmonic-oscillator states with potentially different lengths for the $x,y$ directions to include the case of anisotropic dot,
\seq{
    \phi^\alpha_{n_x,n_y}(x,y)
    = \prod_{\ell=x,y}
      \frac{H_{n_\ell}\!\l(\ell/a^\alpha_{{\rm eff},\ell}\r)}{\sqrt{2^{n_\ell}\, n_\ell!\, \sqrt{\pi}\, a^\alpha_{{\rm eff},\ell}}}\,
      e^{-\frac{1}{2}\l(\ell /a^\alpha_{{\rm eff},\ell}\r)^2},
    \label{eq:fock_darwin}
}
where $H_{n}$ are the Hermite polynomials, $n_x, n_y \geq 0$, and $a^\alpha_{{\rm eff},\ell}$ is the effective oscillator length along direction $\ell$ for band $\alpha$.
An out-of-plane field component $\Bperp$ squeezes the orbitals through the additional magnetic confinement,
\seq{
    \frac{1}{\l(a^\alpha_{{\rm eff},\ell}\r)^4}
    = \frac{1}{\l(a_\ell^{\alpha}\r)^4} + \frac{1}{4\,\ell_B^4},
    \label{eq:magnetic_length}
}
where $a_\ell^{\alpha}$ is the zero-field confinement length used in Eq.~\eqref{eq:Vpar}, and $\ell_B = \sqrt{\hbar/(e\Bperp)}$ is the magnetic length.
At the isotropic point $\varepsilon = 0$, where $a^\alpha_{{\rm eff},x} = a^\alpha_{{\rm eff},y}$, the in-plane eigenstates reorganize into the familiar Fock-Darwin states~\cite{Fock1928,Darwin1931}, eigenstates of the in-plane orbital angular momentum, within each shell of fixed $N = n_x + n_y$.
The quantum numbers $(n_x,n_y)$ in Eq.~\eqref{eq:basis_product} enumerate the in-plane orbital states in order of ascending zero-field energy $\hbar\omega^\alpha_x \l(n_x + \tfrac{1}{2}\r) + \hbar\omega^\alpha_y \l(n_y + \tfrac{1}{2}\r)$.
We include all states up to a maximum shell $N_{\rm max} \geq n_x + n_y$, yielding $N_{\parallel} = (N_{\rm max}+1)(N_{\rm max}+2)/2$ in-plane basis functions per band.
Note that, due to the different effective masses of HHs versus LHs, these basis functions will in general be different via the effective oscillator length.

More specifically, the $N = 0$ ground state is a Gaussian,
\seq{
    \phi^\alpha_{0,0}(x,y)
    = \frac{1}{\sqrt{\pi\, a^\alpha_{{\rm eff},x}\, a^\alpha_{{\rm eff},y}}}\,
      e^{-\frac{1}{2}\l[ \l(x/a^\alpha_{{\rm eff},x}\r)^2 + \l(y/a^\alpha_{{\rm eff},y}\r)^2 \r]},
    \label{eq:s_state}
}
and the first excited shell ($N=1$) consists of the two states with $n_x + n_y = 1$,
\seq{
    \phi^\alpha_{1,0}(x,y)
    &= \frac{\sqrt{2}\, x}{a^\alpha_{{\rm eff},x}}\, \phi^\alpha_{0,0}(x,y), \\
    \phi^\alpha_{0,1}(x,y)
    &= \frac{\sqrt{2}\, y}{a^\alpha_{{\rm eff},y}}\, \phi^\alpha_{0,0}(x,y), 
    \label{eq:p_states}
}
which are the $p$-orbital states ($p_x$ and $p_y$, respectively) central to the qubit encoding described in Sec.~\ref{sec:quadrupole}; at the isotropic point the combinations $\l(\phi^\alpha_{1,0} \pm i\,\phi^\alpha_{0,1}\r)/\sqrt{2}$ carry orbital angular momentum $\pm\hbar$.
Anisotropy, SOC, and Zeeman interactions split the shell into the four levels sketched in Fig.~\ref{fig:main}(b), predominantly composed of $\{\ket{p_x,\!+3/2}, \ket{p_x,\!-3/2}, \ket{p_y,\!+3/2}, \ket{p_y,\!-3/2}\}$ mixtures.

The out-of-plane potential $V_\perp^{\alpha}(z)$ of Eq.~\eqref{eq:Vperp} consists of a linear electric-field term $-eF_z z$ superimposed on the rectangular quantum-well confinement of the Ge layer with potential step $\Delta U_\alpha$, bounded above by the hard wall at $z = d_i$. 
Within each layer (SiGe cap, Ge well, and SiGe buffer) the Schr\"{o}dinger equation reduces to an Airy equation, and the sub-band energies $E^\alpha_{z,k}$ and normalized wavefunctions $\psi^\alpha_k(z)$ are obtained by matching the piecewise Airy solutions through Ben-Daniel--Duke boundary conditions~\cite{BenDaniel1966} at the interfaces.
The explicit construction and the resulting sub-band wavefunctions (Fig.~\ref{fig:airy_wfc}) are given in Appendix~\ref{app:airy}. 
Note that for $F_z$ below a critical value~\cite{Hosseinkhani2020}, the lowest HH sub-bands are localized inside the quantum well, while higher sub-bands and all LH sub-bands increasingly hybridize with the triangular potential at the upper SiGe interface (see Appendix~\ref{app:airy}).

The total orbital basis $\{\Psi^\alpha_{n_x,n_y,k}\} = \{\phi^\alpha_{n_x,n_y}(x,y)\,\psi^\alpha_k(z)\}$ spans both the HH and LH bands, with $N_{\parallel}$ in-plane and $N^{H}_\perp$ and $N^{L}_\perp$ out-of-plane functions for the HH and LH bands, respectively. 
Following the results from Ref.~\cite{Wang2024}, where the authors show that more LH states are needed for convergence of $g$-factors compared to HH states, in general we take $N^{H}_\perp< N^{L}_\perp$. 
The total dimension of the truncated Hilbert space is
\seq{
    N_{\rm tot}
    = 4 N_{\parallel}\bigl(N^{H}_\perp + N^{L}_\perp\bigr),
    \label{eq:Ntot}
}
Throughout this work we use $N_{\rm max} = 8$, corresponding to $N_{\parallel} = 45$ in-plane functions, together with out-of-plane truncations $N^{H}_\perp = 4$ and $N^{L}_\perp = 20$ (see Appendix~\ref{app:airy}), with basis states $|J_z\rangle \otimes |\phi^\alpha_{n_x,n_y}\rangle \otimes |\psi^\alpha_k\rangle$, where the factor of 4 accounts for the four spin-$\tfrac{3}{2}$ projection components $J_z \in \{\pm \tfrac{3}{2}, \pm \tfrac{1}{2}\}$. 
Matrix elements of all in-plane operators (e.g., $x,y,k_x,k_y$) between basis states are evaluated analytically using the harmonic-oscillator ladder algebra along each direction, making the basis computationally efficient. 
Out-of-plane matrix elements (e.g., $z,k_z$) are evaluated numerically on the piecewise Airy representation.

\section{Three-Hole Spin-Orbit Quadrupole Qubit}
\label{sec:quadrupole}

\begin{figure}[t]
    \centering
    \includegraphics[trim={0cm 0cm 9.5cm 0cm}, clip,width=0.5\textwidth]{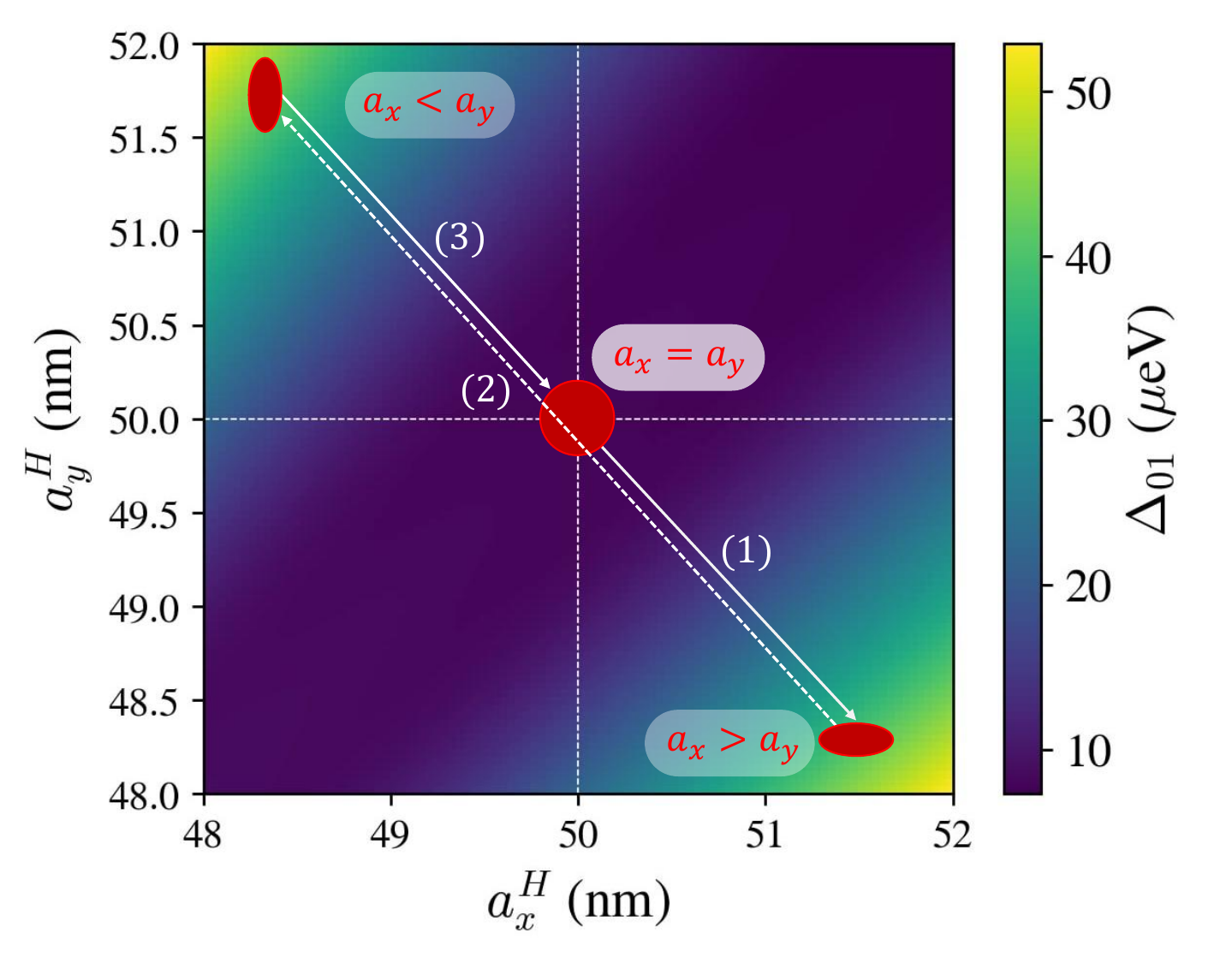}
\caption{
  Qubit level splitting $\Delta_{01}$ as a function of the lateral confinement parameters $(\axH,\ayH)$, with in-plane field $\Bpar=2~$T and $\Bperp=14~$mT, chosen on the sweet-spot locus of Fig.~\ref{fig:dephasing}(b). The labeled points (1)--(3) mark configurations along the antidiagonal cut through the symmetric point $\axH = \ayH$; the level spectrum along this cut is shown in Fig.~\ref{fig:main}(c). The white arrows indicate the Landau-Zener sweep of Sec.~\ref{sec:LZ}, which drives the anisotropy $\varepsilon(t)$ along the antidiagonal, through the avoided crossing at the symmetric point and back.}
\label{fig:2Dspectrum}
\end{figure}

Diagonalizing the Hamiltonian of Eq.~\eqref{eq:Htot} numerically in the truncated basis of Sec.~\ref{ssec:orbital_basis}, we first discard the lowest two spin levels with $s$-orbital flavor assuming they are filled by the first two holes in the shell-filling approximation.
Next, we label its four $p$-shell eigenstates $\{\ket{L}, \ket{0}, \ket{1}, \ket{L'}\}$ in order of increasing energy $E_n$, with $n=L,0,1,L'$.
Here, $\ket{0}$ and $\ket{1}$ will serve as the logical qubit, while $\ket{L}$ and $\ket{L'}$ are considered leakage states. 
These four states are separated from all other levels by the orbital confinement energy, on the order of $1$~meV, and thus define a well-isolated reduced $p$-shell Hamiltonian, written simply as
\seq{
    \Hred(\axH, \ayH) = \sum_{n \in \{L,0,1,L'\}} E_n \ket{n}\!\bra{n}.
    \label{eq:Hp}
}
The in-plane charge distributions of these four states, together with the filled $s$-shell doublet, are shown in Appendix~\ref{app:charge_dist}. 

The dipole matrix elements within the $p$-shell manifold are strongly suppressed: they vanish exactly for pure harmonic $p$ orbitals, and acquire only small values through the HH--LH band mixing and magnetic-field-induced admixtures present in the full LKBP eigenstates.
Consequently, electric-field fluctuations couple states within this manifold predominantly through the quadrupole moment, equivalent to fluctuations in the dot anisotropy.

The spectrum as a function of the lateral confinement anisotropy $\varepsilon$ is shown in Fig.~\ref{fig:main}(c), exhibiting the avoided crossing at $\axH = \ayH \equiv a_0$, with $a_0 = 50\nm$ throughout this work, that underlies the qubit encoding proposed below and the control protocol of Sec.~\ref{sec:LZ}.

Figure~\ref{fig:2Dspectrum} shows the qubit splitting $\Delta_{01} = E_1 - E_0$ as a function of the lateral confinement parameters $(\axH, \ayH)$. 
The splitting varies weakly along the diagonal direction ($\axH = \ayH$), and it is mostly insensitive to perturbations that change the radius of the dot symmetrically.
Along the antidiagonal direction ($\ayH = 2a_0 - \axH$), which changes the anisotropy of the dot at fixed average radius, the splitting remains relatively unchanged in a neighborhood of the symmetric point of about $\pm 0.5\nm$, but changes rapidly outside this region, growing from $\Delta_{01} \approx 10\ueV$ at $(\axH, \ayH) = (50\nm, 50\nm)$ to $\Delta_{01} \approx 50\ueV$ at $(\axH, \ayH) = (48\nm, 52\nm)$.
More explicitly, the antidiagonal cut given by the white arrows of Fig.~\ref{fig:2Dspectrum} corresponds to the spectrum shown in Fig.~\ref{fig:main}(c).
The splitting is thus insensitive to small fluctuations of the anisotropy, providing robustness against noise, as shown in the next section. 
At the same time, inducing a larger anisotropy can be used to drive transitions between the qubit states, which we exploit for all-electrical control in Sec.~\ref{sec:LZ}.
This property constitutes the central motivation for the qubit operating mode proposed in this work.

We propose encoding the logical qubit in the levels $\ket{0}$ and $\ket{1}$, the first and second excited levels of the $p$-shell manifold of the Ge hole dot, operated at the isotropic point $\varepsilon \approx 0$, in analogy with the five-electron Si proposal of Ref.~\cite{Caporaletti2025}. 
At this operating point the qubit splitting is first-order insensitive to fluctuations of the confinement: the anisotropy derivative $\partial\Delta_{01}/\partial\varepsilon$ vanishes at $\varepsilon = 0$, and the size derivative $\partial\Delta_{01}/\partial a_0$ remains small there and over an extended region around the operating point, providing enhanced robustness.
The residual coupling to charge noise that remains at this operating point is quantified in Sec.~\ref{ssec:charge_noise}. 
The result is a qubit whose splitting is robust to charge noise in both the $\axH$ and $\ayH$ directions (alternatively, in the change of anisotropy $\varepsilon$ and average size $a_0$), in contrast to single-axis sweet spots where protection is only one-dimensional in parameter space.

We emphasize that, owing to the strong HH--LH mixing of the Ge valence band and the sweet-spot magnetic field orientation of Sec.~\ref{ssec:charge_noise}, the qubit states at this operating point are coherent mixtures of spin and $p$-orbital character [Fig.~\ref{fig:main}(b)]. 
Consequently, we refer to this system as a \emph{spin-$p$-orbital (SpO) qubit}. 
This parallels the recent proposal of a $p$-orbital--valley (pOv) qubit in silicon of Ref.~\cite{Caporaletti2026}; here, spin in the SpO qubit plays an analogous role to valley in the pOv qubit.
We show the eigenstate composition in terms of the basis states in more detail in Appendix~\ref{app:charge_dist}.
The SpO qubit in Ge is therefore a spin-orbit qubit: it retains the all-electrical controllability of Ge hole spin qubits~\cite{Watzinger2018,Hendrickx2020}, while deriving its noise protection from the quadrupolar orbital structure rather than from the Zeeman splitting alone. 
This distinguishes it from single-hole spin-orbit qubits~\cite{Michal2021,Mauro2025}, where the same spin-orbit coupling that enables driving also exposes the Kramers doublet to charge noise.

\section{Noise Robustness}
\label{sec:noise}

Having established the SpO encoding, we now assess its robustness to two dominant noise mechanisms: quasistatic charge noise and phonon-induced relaxation.

\subsection{Charge-noise characterization and sweet spots}
\label{ssec:charge_noise}

We first evaluate the dephasing time $\tstar$ of the SpO qubit due to quasistatic charge noise, which we model using an ensemble of TLFs randomly distributed near the SiGe/oxide interface of the heterostructure. 
Each TLF represents a trapped charge that bistably switches between two positions separated by a displacement vector $\delta\vec{r}_n$, producing a fluctuating electrostatic potential at the quantum dot. 
The model closely follows the approach of Refs.~\cite{Wang2024,Kepa2023}, adapted to the Ge/SiGe geometry considered here.

We consider $N_{\rm TLF}$ TLFs distributed uniformly in a lateral area $A = 300\,{\rm nm} \times 300\,{\rm nm}$ centered on the quantum dot, at a distance $d_i = 60\nm$ above the top of the quantum well, i.e., at the SiGe/oxide interface $z = d_i$ [Fig.~\ref{fig:main}(a)].
The two charge positions of the $n$th TLF are $\vec{r}_n$ and $\vec{r}_n + \delta\vec{r}_n$, with each displacement vector drawn from an isotropic Gaussian distribution, so that $\sqrt{\langle|\delta\vec{r}_n|^2\rangle} = \delta r$ is fixed for all TLFs. 
The two free parameters of the ensemble are the areal TLF density $\rho$ and the displacement magnitude $\delta r$, which are calibrated in Appendix~\ref{app:TLF} to reproduce a reference shift in the dot ground state chemical potential of $\sigma = 3\ueV$ due to charge-noise fluctuations, inferred from measured $1/f$ charge-noise amplitudes in Ge/SiGe devices~\cite{Lodari2021,Hendrickx2024}. 
More specifically, throughout this work we use $\rho = 5\times10^{10}\,\mathrm{cm}^{-2}$, corresponding to $N_{\rm TLF} = 47$ TLFs in the simulation area $A$, with $\delta r = 0.075\nm$.

The $n$th TLF couples to the hole through the change in its Coulomb potential when the trapped charge is displaced,
\seq{
    \delta V_n = V(\vec{r}_n + \delta\vec{r}_n) - V(\vec{r}_n),
    \label{eq:TLF_potential}
}
where $V(\vec{r})$ is the Coulomb operator obtained from the potential of a point charge at $\vec{r}$. 
We then project $\delta V_n$ onto the $p$-shell eigenbasis of Eq.~\eqref{eq:Hp}, with matrix elements $\mel{i}{\delta V_n}{j}$, $i,j \in \{L,0,1,L'\}$, by integrating numerically using the charge distributions of the eigenstates (see Appendix~\ref{app:charge_dist}). 
In the quasistatic limit, each TLF is frozen in one of its two states on the timescale of a single experimental shot, described by a configuration $\{s_n\} \in \{0,1\}^{N_{\rm TLF}}$, and the perturbed $p$-shell Hamiltonian becomes
\seq{
    \tilde{H}_{p} = \Hred + \sum_{n=1}^{N_{\rm TLF}} s_n\, \delta V_n,
    \label{eq:Hp_perturbed}
}
with $\delta V_n$ understood as its projection onto the $p$-shell subspace.

\begin{figure}[t]
    \centering
    \includegraphics[width=0.45\textwidth]{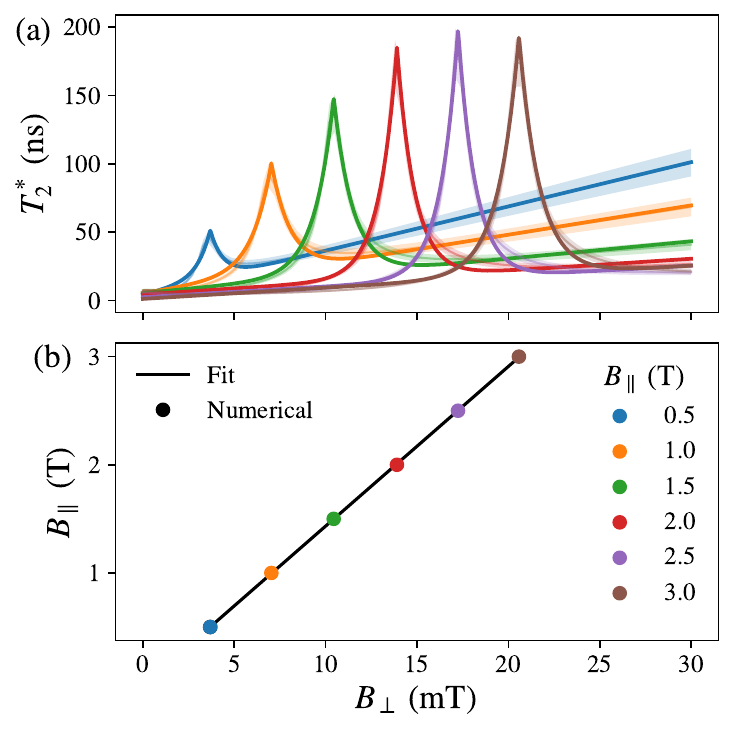}
    \caption{Sweet-spot magnetic field and optimal out-of-plane field component. (a) Average $T_2^*$ as a function of the out-of-plane field component $\Bperp$ for varying in-plane field magnitude $\Bpar$, at the isotropic confinement point $\varepsilon = 0$, $a_0 = 50~$nm. The in-plane field is oriented at $45^\circ$ between the $x$- and $y$-axes. Each data point is averaged over 100 charge-noise configurations, and shaded regions represent the 95\% confidence interval obtained by bootstrapping with replacement. Solid lines are fits to a peaked ansatz with a linear background given by $B_\parallel(B_\perp)=c_0\exp(-c_1 |B_\perp-\Bperp^{\rm opt}|)+c_2 B_\perp$, identifying the optimal out-of-plane field $\Bperp^{\rm opt}$ that maximizes coherence for each $\Bpar$. (b) Fitted sweet-spot values $\Bperp^{\rm opt}$ as a function of $\Bpar$ (circles), together with a linear fit (solid line). The slope yields the optimal magnetic field angle $\theta^*$ between $\bm{B}$ and the $xy$-plane.}
\label{fig:dephasing}
\end{figure}

To leading order in the TLF couplings, the $n$th fluctuator shifts the qubit splitting by $\mel{1}{\delta V_n}{1} - \mel{0}{\delta V_n}{0}$~\cite{Wang2024}. 
However, first-order perturbation theory fails near the sweet spot that constitutes our operating point, and we thus diagonalize the Hamiltonian of Eq.~\eqref{eq:Hp_perturbed} exactly for each TLF configuration.
With the perturbed eigenenergies $\{\tilde{E}_L,\tilde{E}_0,\tilde{E}_1,\tilde{E}_{L'}\}$ obtained from diagonalization, we define the shift due to the ensemble of fluctuators as $\delta E(\{s_n\}) = (\tilde{E}_1-\tilde{E}_0) - \Delta_{01}$, with $\Delta_{01}$ the unperturbed qubit splitting of Sec.~\ref{sec:quadrupole}. 
Assuming equal occupation probability for the two states of each TLF, we compute the standard deviation of $\delta E$ over 1000 randomly sampled TLF state configurations $\{s_n\}$ (out of the $2^{N_{\rm TLF}}$ possible ones), $\sigma_p \equiv \mathrm{std}(\delta E)$.
This procedure is repeated over an ensemble of 100 independent TLF spatial configurations $\{\vec{r}_n, \delta\vec{r}_n\}$.
Lastly, we define the dephasing time using the standard expression derived in the quasistatic noise limit~\cite{Dial2013},
\seq{
    T_2^* = \frac{\hbar\sqrt{2}}{\mathbb{E}[\sigma_p]},
    \label{eq:T2star_sigma}
}
where $\mathbb{E}[\sigma_p]$ is the ensemble average of the energy shift over the spatial configurations.
The confidence intervals are obtained by bootstrapping the ensemble with replacement. 

We apply this framework to the isotropic dot configuration $\varepsilon = 0$ as a function of the magnetic-field components $(\Bpar, \Bperp)$ defined in Sec.~\ref{ssec:hamiltonian}, with the in-plane field angle kept at $45^\circ$ so that $B_x = B_y = \Bpar/\sqrt{2}$. 
Figure~\ref{fig:dephasing} shows the resulting $\tstar$ as a function of $\Bperp$ for several values of $\Bpar$.
For realistic magnetic field amplitudes on the order of 1--2~T, we obtain $T_2^*$ in the range of 100--200~ns, comparable to Ge hole-spin qubits~\cite{Hendrickx2020,Hendrickx2021,Hendrickx2024}.
As shown later, we expect the baseband all-electrical control of the SpO qubit to allow for single and two qubit gate times on the order of $\sim1$--$10$~ns, leading to high quality factors. 

For each $\Bpar$, $\tstar$ exhibits a pronounced peak at an optimal field value $\Bperp^{\rm opt}$, identifying a magnetic-field sweet spot at which the qubit splitting is least sensitive to the dominant charge-noise fluctuations. 
We fit each curve to a peaked ansatz with a linear background to extract $\Bperp^{\rm opt}$.
The linear dependence of $\Bperp^{\rm opt}$ on $\Bpar$ shown in Fig.~\ref{fig:dephasing}(b) reveals that the sweet spot traces a fixed direction in the $(\Bpar, \Bperp)$ plane, corresponding to a well-defined optimal polar angle $\theta^* = \arctan(\Bperp^{\rm opt}/\Bpar)\approx0.4^\circ$ between $\vec{B}$ and the $xy$-plane. 
Practically, $\theta^*$ marks the magnetic field direction for which the qubit splitting is least sensitive to the electric-field fluctuations produced by the TLF ensemble.
We also explored the effects of changing the electric field parameter in the range of $0.5$--$3~$MV/m and found no significant deviations.
Note that this value is somewhat consistent with the results found in Ref.~\cite{Wang2024} for a spin qubit in germanium, where the authors found sweet spots at $\theta\approx0.2^\circ$ for $F_z$ in the range 0.5--2~MV/m.
This is likely not a coincidence: as shown in Appendix~\ref{app:charge_dist}, even though the relative phases for orbital states differ, the charge distributions of $\ket{0}$ and $\ket{1}$ are nearly identical, so the coupling to charge-noise may be dictated predominantly by the spin degree of freedom, rather than by the orbital structure.
If so, the relevant sweet-spot condition is likely set by an effective symmetry between in-plane and out-of-plane dynamics. 
Estimating the $g$-factors from the bare HH states yields $g_\perp = 6\kappa + \frac{27}{2}q \approx 21.4$ and $g_\parallel = 3q \approx 0.2$, and the magnetic field direction at which the in-plane and out-of-plane Zeeman couplings become comparable is $\theta^* \approx \arctan(g_\parallel/g_\perp) \approx 0.5^\circ$, which is in the same order of magnitude as both our results and those of Ref.~\cite{Wang2024}.
Although not a strict derivation, since spin-orbit coupling and HH--LH admixture will modify this picture quantitatively, this analysis offers a simple estimate for the presence of sweet spots and their coarse location in parameter space.


\subsection{Phonon-induced relaxation}
\label{ssec:relaxation}

\begin{figure}[t]
    \centering
    \includegraphics[width=0.45\textwidth]{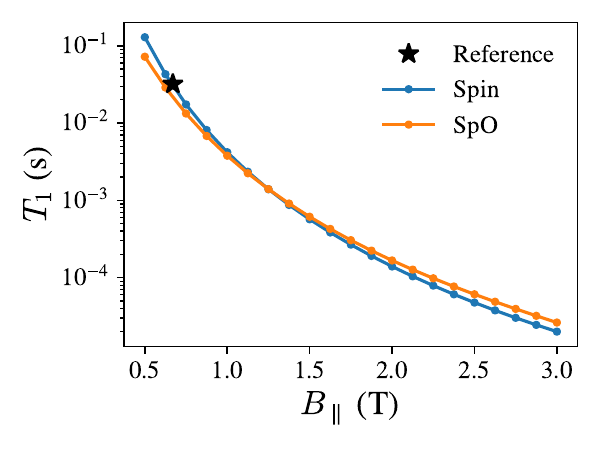}
\caption{Phonon-limited relaxation time $T_1$ of the SpO qubit as a function of in-plane magnetic field $\Bpar$, compared to that of the single-hole spin qubit. The black star indicates the operating field of a reference experimental device~\cite{Hendrickx2024}, showing good agreement with the phonon-induced relaxation model. Both curves are computed using Eq.~\eqref{eq:Gamma_ph} in the dipole approximation.}
\label{fig:relaxation}
\end{figure}

The orbital levels of the SpO qubit are coupled to the phonon bath of the host lattice through the deformation potential.
Since the computational states $\ket{0},\ket{1}$ are admixtures of orbital and spin states, and their splitting is set by orbital confinement energies (tens of $\ueV$) rather than the much smaller Zeeman scale, phonon-induced relaxation can in principle be a practical limitation. 
In this section, we show that this is not the case, as relaxation times in the range of magnetic field considered are much longer than operation times, leading to high quality factors.

To evaluate this quantitatively, we consider one-phonon transitions from the excited qubit state $\ket{1}$ downward, both to the qubit ground state $\ket{0}$ and to the lowest $p$-orbital state $\ket{L}$, with respective transition energies $\Delta_{01}\approx10~\mu$eV and $E_1 - E_L\approx26~\mu$eV.
As mentioned previously, within the shell-filling approximation the lowest two $s$-orbital spin states are occupied by two of the three holes, and consequently relaxation into these fully occupied lower-energy states is considered to be Pauli-blocked.

For a bulk (3D) isotropic phonon bath, the relaxation rate between any two states $\ket{m}$ and $\ket{n}$ ($E_m > E_n$) is given by Fermi's golden rule~\cite{Li2020,Bulaev2005}:
\seq{
    \Gamma_{mn}^{\mathrm{3D}} &= \frac{\omega_{mn}^3}{8\pi^2 \hbar \rho_{\rm Ge}}
    \coth\!\left(\frac{\hbar\omega_{mn}}{2k_B T}\right) \times \\
    &\sum_{\nu \in \{l,\, t_1,\, t_2\}}
    \frac{1}{v_\nu^5}
    \int d\hat{\Omega}
    \bigl|\bra{n} e^{i q_\nu \hat{\mathbf{q}}(\hat{\Omega}) \cdot \mathbf{r}}\, H_\mrm{BP}\bigl[\epsilon_\nu(\hat{\Omega})\bigr] \ket{m}\bigr|^2,
    \label{eq:Gamma_ph}
}
where $\hat{\Omega}=(\theta,\varphi)$ denotes the phonon wavevector direction, $\int d\hat{\Omega} = \int_0^\pi d\theta\, \sin\theta \int_0^{2\pi} d\varphi$ is the solid-angle integral, and the total relaxation rate out of $\ket{1}$ is $\Gamma_{10} + \Gamma_{1L}$.
$\omega_{mn} = (E_m - E_n)/\hbar$ is the angular frequency of the emitted phonon, set by the level splitting between states $\ket{m}$ and $\ket{n}$. 
$\rho_{\rm Ge}=5323~\mathrm{kg/m}^3$ is the mass density of the host material, $T=100~$mK is the considered device temperature and $k_B$ is Boltzmann's constant. 
The $\coth$ factor accounts for stimulated emission and absorption, approaching unity at low temperatures $k_B T \ll \hbar\omega_{mn}$ where spontaneous phonon emission dominates. 
The sum runs over the three acoustic branches, one longitudinal ($l$) and two transverse ($t_1$, $t_2$), with longitudinal and transverse sound velocities $v_l \approx 5400~\mathrm{m/s}$ and $v_t \approx 3200~\mathrm{m/s}$ for Ge~\cite{Li2020}. 
$\hat{\mathbf{q}}(\hat{\Omega}) = (\sin\theta\cos\varphi,\sin\theta\sin\varphi,\cos\theta)$ is the unit phonon wavevector in spherical coordinates, with the resonance condition $v_\nu q_\nu = \omega_{mn}$ selecting only phonons resonant with the transition frequency.
$\epsilon_\nu(\hat{\Omega})$ is the dimensionless $3\times 3$ strain tensor associated with branch $\nu$ at wavevector direction $(\theta,\varphi)$, encoding the polarization of each phonon mode; its explicit construction is given in Appendix~\ref{app:phonon}~\cite{Li2020}. 
Finally, $H_\mrm{BP}[\epsilon_\nu]$ is the deformation-potential perturbation to the LKBP Hamiltonian, i.e., the Bir-Pikus Hamiltonian evaluated at the phonon strain. 
Because the phonon strain in general carries shear components, the diagonal form of Eq.~\eqref{eq:HBP} is supplemented by the shear blocks of the full Bir-Pikus Hamiltonian, also given in Appendix~\ref{app:phonon}. 

The matrix element $\bra{n} e^{iq_\nu\hat{\mathbf{q}}\cdot\mathbf{r}} H_\mrm{BP}[\epsilon_\nu] \ket{m}$ is evaluated numerically at the operating point $(\axH, \ayH)=(50~\mrm{nm},50~\mrm{nm})$. 
To simplify the computation, we use the fact that the size of the dot considered ($a_0\approx 50~$nm) is much smaller than the wavelengths of the resonant phonons, $\lambda_\nu = 2\pi v_\nu/\omega_{mn} \approx 0.4$--$2~\mu\mathrm{m}$ for transition energies of 10--30~$\ueV$, and approximate to first order $e^{iq_\nu\hat{\mathbf{q}}\cdot\mathbf{r}} \approx 1 + iq_\nu\hat{\mathbf{q}}\cdot\mathbf{r}$.
We have confirmed that the first-order results are consistent with a second-order consideration, indicating that this approximation is valid.
The relaxation time of the qubit state $\ket{1}$ is then $T_1 = (\Gamma_{10} + \Gamma_{1L})^{-1}$.

Figure~\ref{fig:relaxation} shows the phonon-limited relaxation time $T_1$ of the SpO qubit as a function of in-plane magnetic field $\Bpar$, alongside that of the single-hole spin qubit for comparison.
Both $T_1$ curves follow similar power-law decay with increasing $\Bpar$. 
Interestingly, the SpO qubit relaxation time is comparable to that of the spin qubit throughout the range of magnetic field considered in this work. 
At the reference experimental field (black star), our estimated spin $T_1$ is of order $10^{-2}$~s, consistent with the reference value. 
In the range of $B_\parallel=1$--$2~$T where $T_2^*$ is long (see Sec.~\ref{ssec:charge_noise}), the relaxation time is $T_1=0.1$--$1~$ms.

As shown in Sec.~\ref{sec:LZ}, this is well above typical gate operation times (on the order of a few ns), confirming that phonon-induced relaxation is not a limiting factor for the SpO qubit under realistic operating conditions.
Lastly, we note that the relaxation rate is strongly dominated by the leakage channel $\ket{1}\rightarrow \ket{L}$, i.e., $\Gamma_{1L}\gg\Gamma_{10}$ and $1/T_1\approx\Gamma_{1L}$. 
Moreover, since these relaxation events take the system out of the computational subspace, they can, provided the leakage is detected, be converted into erasure errors, i.e., errors with known location, which specifically designed quantum error-correcting codes have been shown to tolerate at substantially higher thresholds than Pauli errors~\cite{Grassl1997,Stace2009,Wu2022,Kubica2023}.
This provides further optimism regarding the feasibility of the SpO qubit in Ge as an information processing unit.

\section{Single-Qubit Control}
\label{sec:LZ}

All-electrical control of the SpO qubit requires a mechanism for coherently transferring population between $\ket{0}$ and $\ket{1}$ without microwave drives. 
We exploit the avoided crossing between these two orbital levels [Fig.~\ref{fig:main}(c)] via an LZ sweep of the dot anisotropy $\varepsilon$ at fixed average radius $a_0$, driving the system through the anticrossing and back in a time-symmetric protocol (see white arrows in Fig.~\ref{fig:2Dspectrum}).
The sweep path is parametrized as a truncated Fourier sine series,
\seq{
    \axH(t) = a_0 \l[1 + \varepsilon(t) \r], \quad
    \ayH(t) = a_0 \l[1 - \varepsilon(t) \r],
    \label{eq:path}
}
where the anisotropy is modulated as
\seq{
\label{eq:LZ_anisotropy}
\varepsilon(t) &= \sum_{k=1}^{K} \varepsilon_k \sin\!\left(\frac{2\pi k\, t}{t_f}\right),
}
with the isotropic-point radius $a_0=50~$nm, and $K$ is the number of allowed harmonics.
The optimized parameters are the total sweep duration $t_f$, and the harmonic amplitudes $\varepsilon_k$. 
This parametrization automatically satisfies the boundary conditions $\axH(0) = \axH(t_f/2) = \axH(t_f) = a_0$, ensuring the symmetry of the path, and that the system starts and ends at the sweet spot. 
In terms of the anisotropy parameter of Sec.~\ref{ssec:hamiltonian}, the sweep drives $\varepsilon(t)$ through zero at the anticrossing.
We show in this section that this ansatz can be optimized to provide high-fidelity state transfer within the computational subspace, while simultaneously suppressing leakage.

\begin{figure}[t]
    \centering
    \includegraphics[width=\columnwidth]{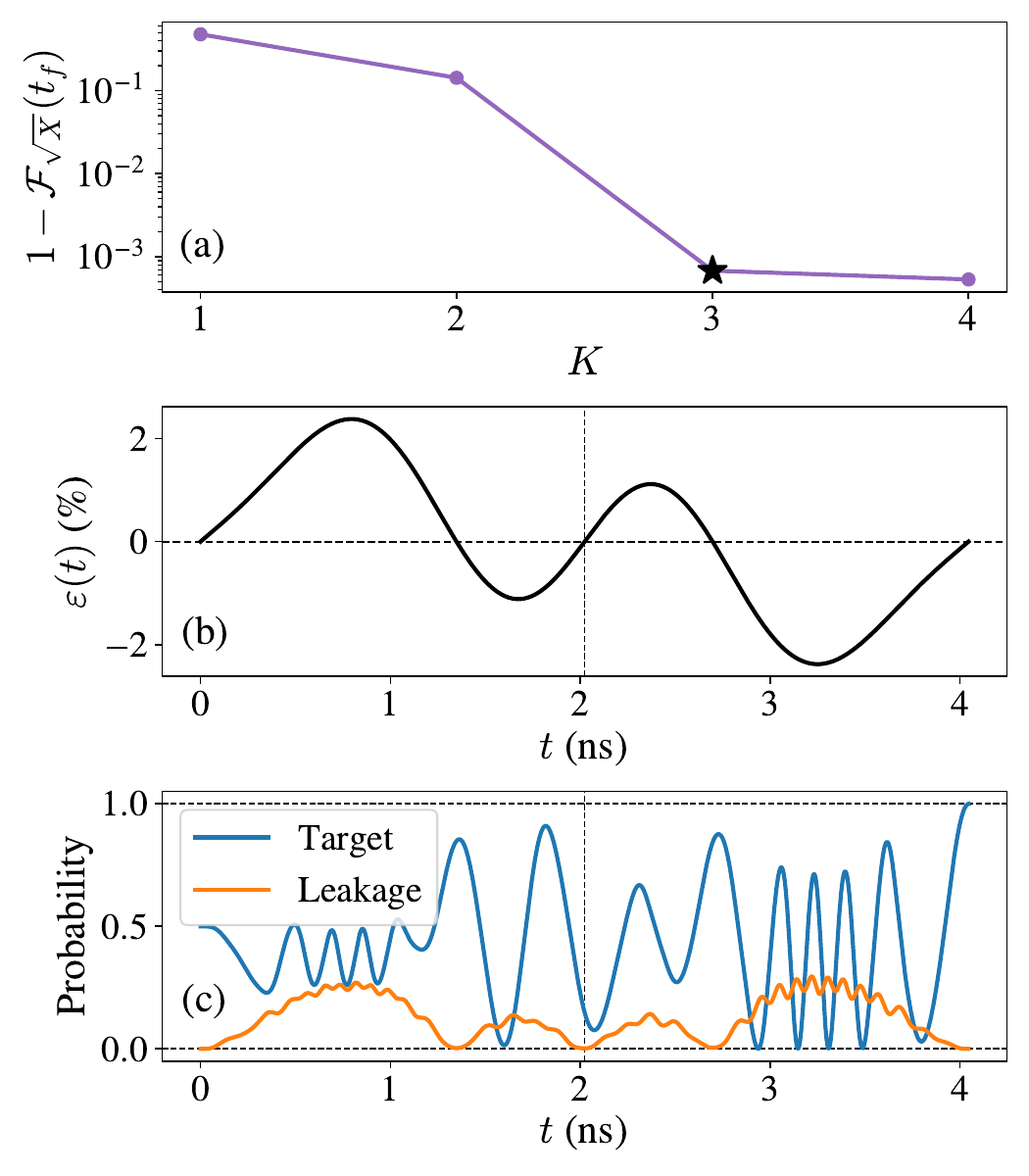}
    \caption{Optimized Landau-Zener state transfer from $\ket{0}$ to $\ket{+_y} \equiv \left(\ket{0} - i\ket{1}\right)/\sqrt{2}$.
    (a)~Gate infidelity $1 - \mathcal{F}_{\sqrt{X}}(t_f)$ as a function of   the number of harmonics $K$ in the sweep parametrization Eq.~\eqref{eq:LZ_anisotropy}. The star marks the solution shown in panels (b) and (c). (b)~Optimized anisotropy $\varepsilon(t)$ [see Eq.~\eqref{eq:LZ_anisotropy}]; the horizontal dashed line marks the sweet-spot value $\varepsilon=0$. (c)~Population dynamics of the target state (blue) and leakage subspace (orange), given by $|\langle \psi(t)| +_y\rangle|^2$ and $|\langle \psi(t)| L\rangle|^2+|\langle \psi(t)| L'\rangle|^2$, respectively, reaching the target state $\ket{+_y}$ with fidelity $\mathcal{F}_{\sqrt{X}} \approx 99.93\%$ at $t_f \approx 4.05$~ns.}
    \label{fig:LZ}
\end{figure}

The four-level Hamiltonian from Eq.~\eqref{eq:Hp}, which we write here as $\Hred[\axH(t), \ayH(t)]$, is evaluated at each time step in the initial-point eigenbasis $\{\ket{L}_0, \ket{0}_0, \ket{1}_0, \ket{L'}_0\}$, defined by diagonalizing $\Hred$ at the starting point $(a_0, a_0)$. 
This ``fixed'' basis approach is used to simplify the numerical calculations, and we have checked that it is accurate to within $1\%$ in the range of anisotropies considered.
Then, the time-ordered unitary propagator
\seq{
    U(t) = \mathcal{T}\exp\!\left(-\frac{i}{\hbar}\int_0^{t} \Hred[\axH(t'), \ayH(t')]\,dt'\right)
    \label{eq:propagator}
}
is computed via piecewise matrix exponentiation on the full four-level system, retaining the $p$-shell leakage channel to the $\ket{L},\ket{L'}$ states, with the time range discretized into uniform time steps $\Delta t=t_f/N_t$ with $N_t=1000$. 

We target the unitary $U_{\sqrt{X}} = \exp\!\left(i\frac{\pi}{4}\sigma_x\right) \oplus \mathbb{1}_{L,L'}$, where $\sigma_x \equiv \ket{0}\!\bra{1}+\ket{1}\!\bra{0}$ acts within the qubit subspace and $\mathbb{1}_{L,L'} \equiv \ket{L}\!\bra{L}+\ket{L'}\!\bra{L'}$ acts as the identity on the leakage states $\ket{L}$ and $\ket{L'}$. Along with the free logical $z$-rotations at the sweet spot, this $\pi/2$ rotation about $x$ provides universal single-qubit control.
The harmonic amplitudes $\{\varepsilon_k\}_{k=1}^K$ and the total time $t_f$ are optimized against the gate fidelity
\seq{
    \mathcal{F}_{\sqrt{X}}(t) = \left| \frac{1}{4} \mathrm{Tr}\!\l[U_{\sqrt{X}}^\dagger\, U(t)\r]\right|^2,
    \label{eq:LZ_fid}
}
by minimizing the loss function
\seq{
    \mathcal{L}_1(t) = 1 - \mathcal{F}_{\sqrt{X}}(t)\exp\l[-(t/\tstar)^2\r]
    \label{eq:L1}
}
at $t = t_f$, where we added a penalty for long gate times compared to the dephasing time $\tstar=200~$ns for $B_\parallel=2~$T. 
Since $t_f \ll \tstar$ for the gates found below, the dephasing factor is close to unity and $\mathcal{L}_1 \approx \mathcal{I}_{\sqrt{X}} \equiv 1 - \mathcal{F}_{\sqrt{X}}$. 
Note that this fidelity metric explicitly penalizes leakage out of the qubit subspace.

The sweep in Eq.~\eqref{eq:LZ_anisotropy} requires a control bandwidth of $\mathrm{BW}\gtrsim  K/t_f$, so we impose the constraint $t_f \geq K/\mathrm{BW}$ on the optimization. 
For a representative bandwidth $\mathrm{BW}=1$~GHz~\cite{Wang2024b} and the maximum harmonic order considered here, i.e., $K=4$, this sets a minimum gate time of $t_f \geq 4$~ns, which we enforce as a lower bound throughout the optimization.
Numerical optimization is performed with the gradient-free Nelder-Mead algorithm, seeded from a single-harmonic ($K=1$) pulse with initial $t_f=4~$ns and initial amplitude $\varepsilon_1=2$~\%.
Then, the solution for the $K$-harmonics case is used as the seed for the $(K+1)$th.

Figure~\ref{fig:LZ}(a) shows the gate infidelity $\mathcal{I}_{\sqrt{X}}$ as a function of the number of harmonics $K$. 
The infidelity decreases systematically with $K$, demonstrating that higher harmonics provide additional freedom to shape the sweep trajectory and are able to undo the diabatic excitation of the $\ket{L}$ and $\ket{L'}$ levels. 
The infidelity reaches $\mathcal{I}_{\sqrt{X}} \approx 4\cdot 10^{-4}$ at $K = 3$, corresponding to a gate fidelity of $\mathcal{F}_{\sqrt{X}} \approx 99.93\%$ with a total gate time of $t_f = 4.05$~ns, which as mentioned earlier is much shorter than estimated $T_2^*$ times. 

Figures~\ref{fig:LZ}(b) and (c) show the optimized sweep trajectory and the resulting population dynamics. 
The confinement parameter $\axH(t)$ departs from the sweet spot, traverses the avoided crossing at $\axH = a_0$ at $t = t_f/2$, and returns to $a_0$ at $t = t_f$. 
The population dynamics in panel (c), with $\ket{\psi(t)} = U(t)\ket{0}$ and target state $\ket{\psi(t_f)}=\ket{+_y} \equiv U_{\sqrt{X}}\ket{0}$, reflect the complexity of the multi-level avoided-crossing structure: significant transient population is transferred to all four levels during the sweep, but the protocol steers the system back to maximize population of $\ket{+_y}$ at $t = t_f$. 
Population out of the qubit subspace at the end of the protocol is rendered negligible, confirming that the gate fidelity is not limited by leakage to the $\ket{L},\ket{L'}$ states.

\section{Qubit-Qubit Interactions and Two-Qubit Gates}
\label{sec:qubit-qubit}

In this section, we address two qubits interacting via Coulomb interaction, with the goal of designing two-qubit entangling gates.
We consider two identical SpO qubits in dots $A$ and $B$ separated by a center-to-center distance $R$, with non-overlapping charge distributions. 
The electrostatic interaction energy between the two dots is
\seq{
V_{AB} = e^2 F_c \int \frac{\rho_A(\vec{r})\,\rho_B(\vec{r}')}{|\vec{r}-\vec{r}'|} d^3r\, d^3r',
\label{eq:UAB_coulomb}
}
where $F_c = (4\pi\epsilon_0\epsilon_r)^{-1}$ and $\epsilon_r \approx 14.67$ is the relative permittivity of Ge~\cite{Wang2024}. 
Next, we expand $|\vec{r}-\vec{r}'|^{-1}$ in a Taylor series about the dot centers~\cite{Jackson1999}, retaining terms through fourth order in the dot coordinates.
The monopole term contributes a state-independent energy offset $e^2 F_c / R$ that plays no role in the qubit-qubit coupling. 
The dipole matrix elements are suppressed due to the $p$-orbital nature of the charge distributions, shown explicitly in Appendix~\ref{app:charge_dist}.
The quadrupole-monopole terms ($\propto R^{-3}$) likewise contribute single-qubit energy shifts that can be calibrated out.

Thus, the leading entangling inter-dot coupling arises at the quadrupole-quadrupole level, where
\seq{
V_{AB}^{(2,2)}(R) = \frac{e^2 F_c}{4} \sum_{i,j,k,l} T_{ijkl}(R)\, Q_{ij}^A \otimes Q_{kl}^B,
\label{eq:UQQ}
}
where $T_{ijkl}(R) = \partial_i\partial_j\partial_k\partial_l R^{-1}$ is the interaction tensor evaluated at the inter-dot separation $\vec{R}$, scaling as $R^{-5}$. 
Here, we consider dots separated along the $x$ direction, i.e., $\vec{R}\parallel \hat{x}$.
The quadrupole moment operators $Q_{ij}^{A/B}$ are promoted to operators in the single-dot orbital basis by quantizing the charge density as $\rho_{A/B}(\vec{r}) \to \sum_{n,m} \psi_n^*(\vec{r})\psi_m(\vec{r})\ket{n}\bra{m}$, giving rise to matrix elements $[Q_{ij}^{A/B}]_{nm} = \int \psi_n^*(\vec{r})\, r_i r_j\, \psi_m(\vec{r})\, d^3r$ computed numerically from the LKBP eigenstates for the effective single-hole wavefunctions. 
Note that these quadrupole moments are centered at the corresponding dot centers, taken as $\vec{R}_A=-\vec{R}/2$ and $\vec{R}_B=\vec{R}/2$. 
Promoting these to the two-dot Hilbert space gives the interaction matrix
\seq{
V_{AB}^{(2,2)} = \sum_{n,m,p,q} V_{nmpq} \, \ket{n, p}\bra{m, q},
\label{eq:UAB_general}
}
where $\ket{n, p} \equiv \ket{n}_A \otimes \ket{p}_B$, with $n,m \in \{L,0,1,L'\}$ indexing dot-$A$ states and $p,q \in \{L,0,1,L'\}$ indexing dot-$B$ states. 
The coefficients $V_{nmpq} = \frac{e^2 F_c}{4}\sum_{ijkl} T_{ijkl} [Q_{ij}^A]_{nm} [Q_{kl}^B]_{pq}$ are the matrix elements of Eq.~\eqref{eq:UQQ}. 

The full two-dot Hamiltonian at separation $R$ is then
\seq{
H_{16}(R) = \Hred \otimes \id + \id \otimes \Hred + V_{AB}^{(2,2)}(R),
\label{eq:H2Q}
}
acting on the $16$-dimensional product space of the two four-level dots. 
The two-qubit computational-subspace states composed of linear combinations of $S_4=\{\ket{0,0},\ket{1,0},\ket{0,1},\ket{1,1}\}$ are clustered closely in energy, and are accompanied by a nearby leakage pair, composed of essentially $(\ket{L,L'} \pm \ket{L',L})/\sqrt{2}$.
The energy spectrum of $H_{16}(R)$ for these six states as a function of $R$ is shown in Appendix~\ref{app:2hole}. 
As shown below, these leakage states remain essentially unpopulated during the adiabatic shuttle proposed here.

\begin{figure}[t]
\centering
\includegraphics[width=0.9\columnwidth]{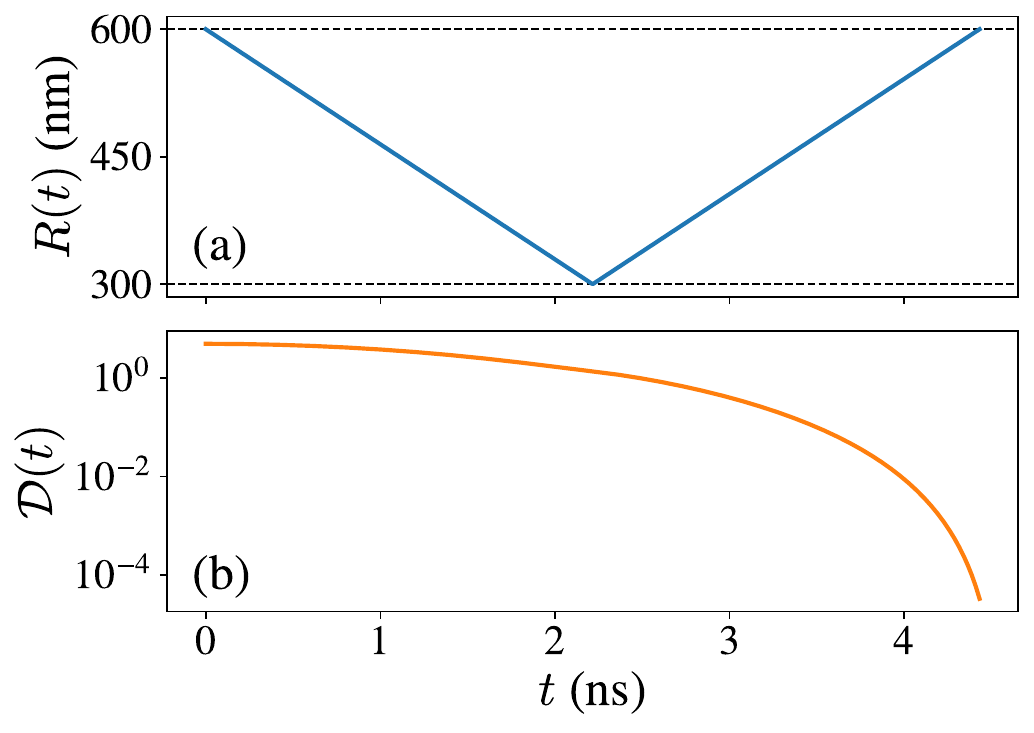}
\caption{Optimized adiabatic two-qubit entangling gate for a constant velocity shuttling profile. (a) Optimized inter-dot separation $R(t)$ between $R_{\max}=600$ nm and $R_{\min}=300$ nm. (b) Invariant distance $\mathcal{D}(t)$ [see Eq.~\eqref{eq:L2}], evaluated on the partially accumulated qubit-subspace propagator $U_4(t)$: starting from the identity class [$(G_1,G_2)=(1,3)$], $\mathcal{D}(t)$ decreases by several orders of magnitude and reaches the target local-equivalence class of $U_{ZZ}=\exp(i\pi/4\,ZZ)$ [$(G_1,G_2)=(0,1)$] at $t_f=4.43$~ns.}
\label{fig:2Q_Makhlin_opt}
\end{figure}

We realize a two-qubit entangling gate by adiabatically shuttling the dots, and thus changing the inter-dot separation $R(t)$ from its idle value $R_\mrm{max}$ down to a minimum separation $R_\mrm{min}$ and back, so that the strong quadrupole-quadrupole Coulomb coupling at small $R$ accumulates an entangling phase. 
For sufficiently short inter-dot distances, the overlap of neighboring hole wavefunctions becomes non-negligible, and one must worry about exchange interaction; thus, we choose the shortest distance to be $R_\mrm{min} = 300~$nm so this effect can be neglected.
Meanwhile, the $R^{-5}$ scaling of Eq.~\eqref{eq:UQQ} guarantees that the residual coupling at the larger idle separation is suppressed by a factor $(R_{\max}/R_{\min})^5$ relative to the gate configuration; here we fix $R_\mrm{max}=600~$nm.

In addition, during the shuttle both dots are operated slightly away from the isotropic point, with a small confinement anisotropy $\varepsilon = 0.22\%$ that enhances the quadrupole-quadrupole coupling. 
At this value of anisotropy, the Coulomb interaction is approximately 2.3 times stronger than at the isotropic point $\varepsilon=0$ (as quantified by the root-mean-squared value of the 16 eigenvalues compared to the non-interacting case at $R_\mrm{min}$), with a reduction of only 40~\% in $T_2^*$. 
As can be seen in Appendix~\ref{app:T2-vs-anisotropy} for $T_2^*$ vs $\varepsilon$, $T_2^*$ decreases exponentially for $\varepsilon\gtrsim 0.22\%$, and so this value becomes the optimal operation point that maximizes the coupling strength while retaining protection to charge-noise.
Single-qubit operation and idling take place at the isotropic point $\axH = \ayH = a_0$, and the dots are squeezed only for the duration of the two-qubit gate. 

The full $16\times16$ two-dot Hamiltonian, which we write $H_{16}[R(t)]$ from Eq.~\eqref{eq:H2Q}, is propagated exactly at each instant via piecewise-exact diagonalization, with no adiabatic approximation imposed on the dynamics, 
\seq{
U_{16}(t)=\mathcal{T} \exp\l( -\frac{i}{\hbar}\int_0^{t} H_{16}[R(t')]\,dt' \r).
}
The qubit-subspace propagator $U_4(t)$ is extracted from the full propagator through its matrix elements on the computational subspace $S_4$, i.e., $[U_4(t)]_{ij} = \mel{i}{U_{16}(t)}{j}$ with $i,j \in S_4$.
Note that $U_4(t)$ is in general not unitary.
In order to minimize any population coherently transferred out of $S_4$, we include a penalty on unitarity loss $\|U_4^\dagger U_4 - \id\|$. 
In the language of quantum channels, $U_4$ is the Kraus operator of the leakage-free branch of the closed two-dot evolution; the complementary Kraus operators map into the leakage sector and are negligible for the optimized gates found below.

Next, to optimize the entangling gate independently of any particular single-qubit frame choice, we characterize $U_{4}(t)$ using the Makhlin local invariants~\cite{Makhlin2002,Zhang2003,Pal2015}, which classify two-qubit gates up to local (single-qubit) unitary equivalence, i.e., $U_1\sim U_2$ if $U_1=(A_1\otimes B_1)\,U_2\,(A_2\otimes B_2)$ for single-qubit unitaries $A_i,B_i$. 
Working in the magic (Bell) basis~\cite{Makhlin2002}, defined by the transformation
\seq{
    \mathcal{Q} = \frac{1}{\sqrt{2}}
    \begin{pmatrix}
        1 & 0 & 0 & i \\
        0 & i & 1 & 0 \\
        0 & i & -1 & 0 \\
        1 & 0 & 0 & -i
    \end{pmatrix},
    \label{eq:magic_basis}
}
whose columns are the maximally entangled Bell states expressed in the computational basis $S_4$, we define $m = \mathcal{Q}^\dagger U_4\, \mathcal{Q}$ and construct
\seq{
G_1 &= \frac{\Tr\l[m^Tm\r]^2}{16\det U_4}, \\
G_2 &= \frac{\Tr\l[m^Tm\r]^2 - \Tr\l[(m^Tm)^2\r]}{4\det U_4}.
\label{eq:makhlin}
}
The pair $(G_1,G_2)$ is invariant under local unitary conjugation and uniquely labels the local-equivalence class of $U_4$.
For example, the entangling gate $U_{ZZ}=\exp(i\pi/4\,ZZ)$, being locally equivalent to the controlled-$Z$ gate, corresponds to $(G_1,G_2)=(0,1)$, while the identity corresponds to $(G_1,G_2)=(1,3)$. 

We use $(G_1,G_2)$ directly in the optimization cost. 
Defining the invariant distance to the target class, $\mathcal{D}(t) = \l|G_1(t)\r|^2+\l|G_2(t)-1\r|^2$, we minimize the loss function
\seq{
    \mathcal{L}_2(t) &= 1 - \l[1 - \mathcal{D}(t)\r]e^{-(t/\tstar)^2} \\
    &+ \l\|U_4^\dagger(t)\, U_4(t) - \id\r\|^2
    \label{eq:L2}
}
at $t = t_f$, rather than the trace fidelity against a fixed target unitary.
In addition, as in Sec.~\ref{sec:LZ}, the dephasing factor penalizes gate times long compared to $\tstar$, while the last term (a Frobenius norm) penalizes leakage out of $S_4$ through the loss of unitarity of $U_4$. 
Here, we use $\tstar\approx100$~ns of Sec.~\ref{ssec:charge_noise}, corresponding to the two-qubit operating field $\Bpar = 1$~T.

As in Sec.~\ref{sec:LZ}, the shuttle waveform is bandwidth-limited.
Assuming a control bandwidth of $\mathrm{BW}=1$~GHz~\cite{Wang2024b} allows an inter-dot shuttle of 150~nm~\cite{VanRiggelen2024} in 1~ns, we estimate the maximum shuttling speed allowed by the control electronics to be $v_\mrm{max}\approx150$~m/s.
Then, a round trip over $\Delta R = R_\mrm{max}-R_\mrm{min}$ at constant speed requires $t_f \geq 2\Delta R/v_\mrm{max}$, which for our parameters becomes $t_f\geq4$~ns. 
For comparison, high fidelity spin shuttling at speeds of $\sim50~$m/s has recently been demonstrated experimentally in Ref.~\cite{VanRiggelen2024}. 
We remark that, although we expect higher shuttling speeds to be achievable soon, it is also possible to optimize the entangling gates with a longer minimum shuttling time.

Next, we parametrize the shuttle shape with a triangular (constant-velocity) profile,
\seq{
    R(t) = 
    \begin{cases}
        R_\mrm{max} - v\,t, & t \leq t_f/2, \\
        R_\mrm{min} + v\,(t-t_f/2), & t > t_f/2,
    \end{cases}
    \label{eq:shuttle_ansatz}
}
with constant speed $v = 2\Delta R/t_f$.
The optimization then reduces to a low-dimensional search, where the total gate time $t_f$ is the only free parameter, with $R_{\min}$, $R_{\max}$ fixed and the round-trip shape otherwise fully specified by symmetry.
Figure~\ref{fig:2Q_Makhlin_opt} shows the optimized profile and the time evolution of the invariant distance $\mathcal{D}(t)$, which decreases by several orders of magnitude as $U_4(t)$ evolves from the identity class to the target class by $t = t_f$, with $t_f = 4.43$~ns, with final invariants satisfying $\mathcal{D}(t_f) \approx 3\cdot 10^{-5}$. 
As in the single-qubit control case, the two-qubit gate time sits comfortably below $\tstar$. 
Since the shuttle-based scheme is not tied to a particular target frame, more specific gates are equally accessible by direct optimization: in Appendix~\ref{app:ZZ} we optimize the same shuttle against the fixed target unitary $U_{ZZ}$ using a smooth flat-top profile, reaching a coherent gate fidelity $\mathcal{F}_{ZZ}(t_f) \approx 1$.

\section{Discussion and Conclusion}
\label{sec:conclusion}

In this work, we have introduced a novel framework for noise protected spin-orbit hole qubits in planar Ge/SiGe.
By filling the quantum dot with three holes, we obtain a charge distribution with suppressed dipole moment, and which couples to electric fields via the quadrupole moment.
This constitutes the Ge version of the pO encoding originally proposed for the five-electron $p$-like valence states of a Si quantum dot~\cite{Caporaletti2025}, which couple to charge noise through the quadrupole moment. 
The present work translates this encoding to planar Ge holes, with three qualitative differences. 
First, the absence of valley degeneracy in the Ge valence band removes the valley-orbit physics that complicates the $p$-shell spectrum in Si. 
Second, the strong intrinsic spin-orbit coupling makes the Ge SpO qubit a spin-orbit qubit: the logical states carry mixed spin and orbital character, which introduces the magnetic-field-orientation sweet spot of Sec.~\ref{ssec:charge_noise} as an additional protection axis absent in the electron system.
However, this is at the price of a spectrum that must be characterized within the full LKBP model rather than a single-band picture, and consists of 4 bands with a qubit hosted not in the lowest two states. 
Third, we quantify the noise protection against a microscopically calibrated TLF ensemble and the relaxation against a phonon bath, showing that the SpO qubit inherits the phonon-limited $T_1$ of the hole spin qubit while being charge-noise resilient and operable via all-electrical baseband control.

We have explored the parameter space in magnetic field and electrostatic confinement to find operation points giving rise to sweet spots where the qubit splitting is first-order insensitive to fluctuations in the electrostatic confinement parameters, i.e.,  $\partial\Delta_{01}/\partial a^H_j=0$ for $j=x,y$. 
In this configuration, we have confirmed numerically that the coupling to charge noise is suppressed using a TLF model. 
Simultaneously, this operation mode allows for fast qubit control via a LZ sweep of the anisotropy of the dot. 
The LZ protocol provides an all-electrical path to logical qubit control, and the quadrupole-quadrupole Coulomb interaction enables two-qubit entangling gates operated by adiabatic shuttling.

A natural question is whether the SpO qubit proposal can be adapted to Si holes. 
The SpO mechanism relies on strong spin-orbit coupling to admix spin into the $p$-orbital manifold.
Given the weaker SOC of Si holes~\cite{Winkler2003}, it is unclear whether the same mechanism would give rise to a noise protected logical space, as was shown in this work for Ge. 
Si holes also present additional differences: their in-plane effective mass ($0.216\,m_0$) is roughly four times larger than in Ge ($0.057\,m_0$), thus requiring smaller dot sizes, and their HH--LH splitting ($\sim5$~meV, related to Si's smaller split-off gap, $\Delta_{\rm SO}\approx44$~meV) is at least an order of magnitude smaller than in Ge, leading to much stronger HH--LH admixture, and the need to retain all 6 bands for accurate modeling~\cite{Wang2024c}. 
We therefore expect Ge to remain the more favorable host for the SpO encoding, where noise protection and control rely on coherent spin-orbit admixture. 
However, it is worth noting that the original encoding into the p-orbital manifold proposed in Ref.~\cite{Caporaletti2025} (i.e., the pO qubit) consists of a purely orbital encoding, and does not require spin-orbit admixture.
In addition, as opposed to the original pO proposal in conduction band electrons, Si holes do not present the additional complexity of valley degeneracy~\cite{Winkler2003}.
Consequently, the pO qubit encoding may still be attractive for Si holes.
A dedicated, quantitative study of $p$-orbital encodings in Si holes is left for future work.

Regarding experimental feasibility, the protocols proposed here rely on control knobs that are all available in current devices. 
State-of-the-art timing precisions of $\approx4~$ps~\cite{Wang2024b} and remarkable recent experimental progress provides optimism regarding feasibility of practical implementation of this protocol.
Gate timescales in the order of $t_f = 4$~ns can be achieved for both the single-qubit $\pi/2$-pulse and two-qubit entangling gates, and are limited by the bandwidth of control electronics rather than any fundamental considerations.
The flatness of $\Delta_{01}$ along the antidiagonal of Fig.~\ref{fig:2Dspectrum} translates the sweet-spot condition into a tolerance on the confinement anisotropy of $|\varepsilon| \lesssim 0.5\%$ for $B_\parallel=2~$T (see App.~\ref{app:T2-vs-anisotropy} for $T_2^*$ vs $\varepsilon$ in the $B_\parallel=1~$T case).
These gate times are roughly two orders of magnitude below the charge-noise induced $T_2^* \approx 100$--$200$~ns, and more than four orders of magnitude below the phonon-limited $T_1 = 0.1$--$1$~ms, for operating magnetic fields in the range of $\Bpar = 1$--$2$~T. 
The adiabatic shuttling required for the two-qubit gate is compatible with the gate-controlled dot displacement and coherent shuttling recently demonstrated in Ge hole devices~\cite{VanRiggelen2024,Wang2024b}. 
A quantitative error budget for these gates under realistic pulse distortion and residual noise is left for future work. 
In summary, combining sweet-spot operation, $p$-orbital encoding, LZ control, and quadrupole-mediated coupling, establishes the SpO Ge qubit as a promising candidate with realistic hardware requirements and high quality factors, without requiring micromagnets or complex pulse sequences.

\begin{acknowledgments}
This work was supported by the Army Research Office (ARO) under Grant No. W911NF-23-1-0115. 
The authors thank John H. Caporaletti, Dr. Adrian B. Culver, Dr. David W. Kanaar and Dr. Mark F. Gyure for useful discussions.
\end{acknowledgments}

\emph{Code availability.} All numerical results in this work were produced with our open-source LKBP quantum-dot simulation code, available at \url{https://github.com/y-oda2/ge-qd-lkbp}.

\appendix

\section{Spin-3/2 angular momentum matrices}
\label{app:Jmatrices}

The spin-$\tfrac{3}{2}$ angular momentum matrices $J_\ell$ ($\ell=x,y,z$) entering the Zeeman Hamiltonian of Eq.~\eqref{eq:Zeeman} are given explicitly, in the same basis $\{\ket{\tfrac{3}{2},\tfrac{3}{2}}, \ket{\tfrac{3}{2},-\tfrac{3}{2}}, \ket{\tfrac{3}{2},\tfrac{1}{2}}, \ket{\tfrac{3}{2},-\tfrac{1}{2}}\}$ used for $H_\mathrm{Z}$ in Eq.~\eqref{eq:Zeeman}, by~\cite{Winkler2003}
\seq{
J_x &= \frac{1}{2}
\begin{pmatrix}
0 & 0 & \sqrt{3} & 0 \\
0 & 0 & 0 & \sqrt{3} \\
\sqrt{3} & 0 & 0 & 2 \\
0 & \sqrt{3} & 2 & 0
\end{pmatrix}, \\
J_y &= \frac{i}{2}
\begin{pmatrix}
0 & 0 & -\sqrt{3} & 0 \\
0 & 0 & 0 & \sqrt{3} \\
\sqrt{3} & 0 & 0 & -2 \\
0 & -\sqrt{3} & 2 & 0
\end{pmatrix}, \\
J_z &= \frac{1}{2}
\begin{pmatrix}
3 & 0 & 0 & 0 \\
0 & -3 & 0 & 0 \\
0 & 0 & 1 & 0 \\
0 & 0 & 0 & -1
\end{pmatrix}.
\label{eq:Jmatrices}
}
These satisfy the standard angular-momentum commutation relations $[J_i,J_j]=i\,\epsilon_{ijk}J_k$, and enter the anisotropic Zeeman term $\mathcal{J}_\ell = \kappa J_\ell + q J_\ell^3$ of Eq.~\eqref{eq:Zeeman} directly in this basis.

\section{Out-of-plane sub-band wavefunctions}
\label{app:airy}

\begin{figure}[t!]
    \centering
    \includegraphics[trim={0 0cm 0cm 0}, clip,width=0.48\textwidth]{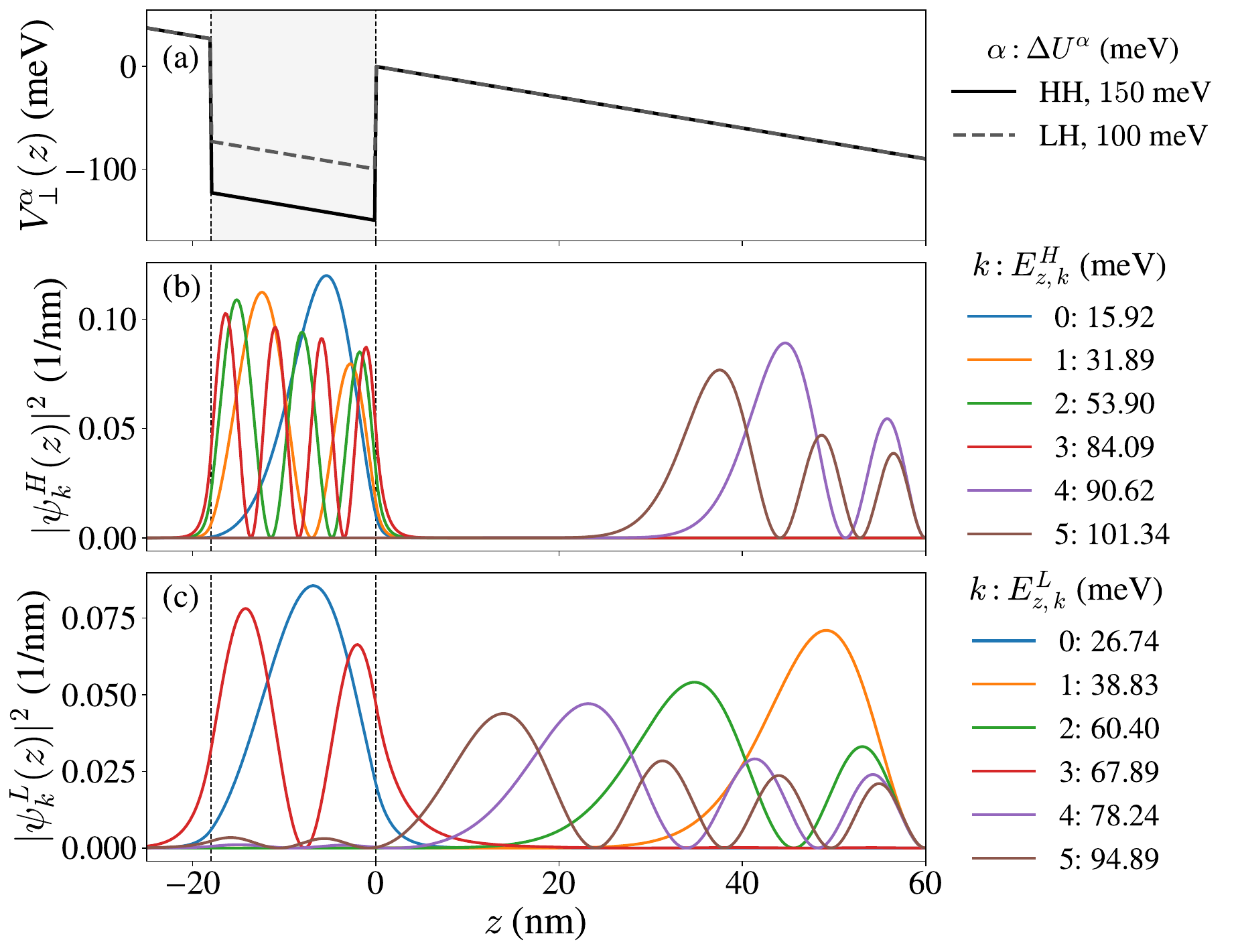}
    \caption{(a) Vertical potential for HH (solid line) and LH (dashed line). (b,c) Out-of-plane sub-band probability densities $|\psi^\alpha_k(z)|^2$ of the six lowest (b) HH and (c) LH sub-bands, at $F_z = 1.5$~MV/m, with the corresponding sub-band energies $E^\alpha_{z,k}$ listed in the legends. Vertical dashed lines mark the Ge quantum well $-d_w < z < 0$. Only the lowest sub-bands are localized inside the well; higher-lying sub-bands hybridize with the triangular potential formed in the SiGe cap by the vertical field and localize toward the hard wall at $z = d_i$.}
\label{fig:airy_wfc}
\end{figure}

The out-of-plane potential $V_\perp^{\alpha}(z)$ of Eq.~\eqref{eq:Vperp} is piecewise-defined. 
In each region (SiGe cap $0 < z < d_i$, Ge well $-d_w < z < 0$, and SiGe buffer $z < -d_w$) the Schr\"{o}dinger equation reduces to an Airy equation of the form
\seq{
    -\frac{\hbar^2}{2m^\alpha_\perp}\frac{d^2\psi}{dz^2}
    + (-eF_z z + V_r)\psi = E\psi,
    \label{eq:airy_ode}
}
where $V_r$ is the (constant) band offset in each region $r$, satisfying $|V_\mrm{SiGe}-V_\mrm{Ge}|=\Delta U_\alpha$, and $m^{H/L}_\perp=m_0/(\gamma_1\mp2\gamma_2)$ is the out-of-plane effective mass, both depending on the sub-band $\alpha=H,L$. 
The general solution in region $r$ is a linear combination of the two independent Airy functions,
\seq{
    \psi^\alpha_k(z)\big|_r
    = c^{\alpha,r}_{k,1}\,\mathrm{Ai}(\zeta^\alpha_r(z))
    + c^{\alpha,r}_{k,2}\,\mathrm{Bi}(\zeta^\alpha_r(z)),
    \label{eq:airy_solution}
}
where the dimensionless argument is
\seq{
    \zeta^\alpha_r(z)
    &= -\frac{z}{\ell^\alpha_r} + \frac{E - V_{r}}{E^\alpha_{{\rm tri},r}}, \\
    \ell^\alpha_r
    &= \left(\frac{\hbar^2}{2m^\alpha_{\perp,r}\,eF_z}\right)^{1/3},
    \label{eq:airy_argument}
}
with $E^\alpha_{{\rm tri},r} = \hbar^2/(2m^\alpha_{\perp,r} (\ell^\alpha_r)^2)$ the triangular-potential energy scale. 
The coefficients $c^{\alpha,r}_{k,1}$ and $c^{\alpha,r}_{k,2}$ in each region, together with the sub-band energies $E^\alpha_{z,k}$, are determined by imposing the Ben-Daniel--Duke boundary conditions~\cite{BenDaniel1966} at each interface $z = 0$ and $z = -d_w$,
\seq{
    \psi^\alpha_k\big|_{r}(z_{\rm int})
    &= \psi^\alpha_k\big|_{r'}(z_{\rm int}),
    \\
    \frac{1}{m^\alpha_{\perp,r}}
    \partial_z\psi^\alpha_k\big|_{r}(z_{\rm int})
    &= \frac{1}{m^\alpha_{\perp,r'}}
    \partial_z\psi^\alpha_k\big|_{r'}(z_{\rm int}),
    \label{eq:bc_continuity}
}
where $(r,r')$ denotes a pair of adjacent regions meeting at $z_{\rm int}$, supplemented by the hard-wall condition $\psi^\alpha_k(d_i) = 0$ at the oxide interface. 
In the SiGe buffer the wavefunction must decay as $z \to -\infty$, which requires $c^{\alpha,r}_{k,2} = 0$ there, since $\zeta^\alpha_r \to +\infty$ as $z \to -\infty$ and $\mathrm{Bi}(\zeta) \to \infty$ as $\zeta \to +\infty$. In the limit where the effective masses in the Ge well and SiGe barriers are taken to be equal, $m^\alpha_{\perp,\rm Ge} \approx m^\alpha_{\perp,\rm SiGe}$, the boundary conditions Eq.~\eqref{eq:bc_continuity} reduce to simple continuity of $\psi$ and $\partial_z\psi$, and the quantization condition for the sub-band energies $E^\alpha_{z,k}$ is equivalent to finding roots of a transcendental equation involving products of Airy functions. 
These 5 equations define a $5\times5$ matrix $M(E_{z,k}^\alpha)$ with numerical entries, and the roots are satisfied when $\det[M(E_{z,k}^\alpha)]=0$, to ensure that a non-trivial solution to the system of equations exists.
These roots are found numerically using a combination of root isolation on a fine energy grid and a bisection refinement step to ensure no sub-bands are missed.

The resulting sub-band wavefunctions $\psi^\alpha_k(z)$ are normalized, $\int_{-\infty}^{\infty}|\psi^\alpha_k(z)|^2\,dz = 1$, and are orthogonal for different $k$ within the same band.
Figure~\ref{fig:airy_wfc} shows the six lowest sub-bands of each band at the operating field: the lowest HH sub-bands are confined to the Ge well, while higher-lying HH and LH sub-bands increasingly localize in the triangular potential of the SiGe cap.
The low-energy states of the accumulated hole are predominantly of HH character, and thus reside mostly in the well. 
However, some excited LH states are found to be also localized in the well (see $k=3$ in Fig.~\ref{fig:airy_wfc}(c)), and thus contribute to the low-energy $p$-shell physics.
Therefore, even though only a few HH states are sufficient for convergence, a larger number of LH states are needed, and the out-of-plane truncations $N^{H}_\perp = 4$ and $N^{L}_\perp = 20$ are chosen accordingly, after checking for convergence.

\section{In-plane charge distribution of wavefunctions}
\label{app:charge_dist}

\begin{figure}[t]
    \centering
    \includegraphics[trim={0 0cm 0cm 0}, clip,width=0.48\textwidth]{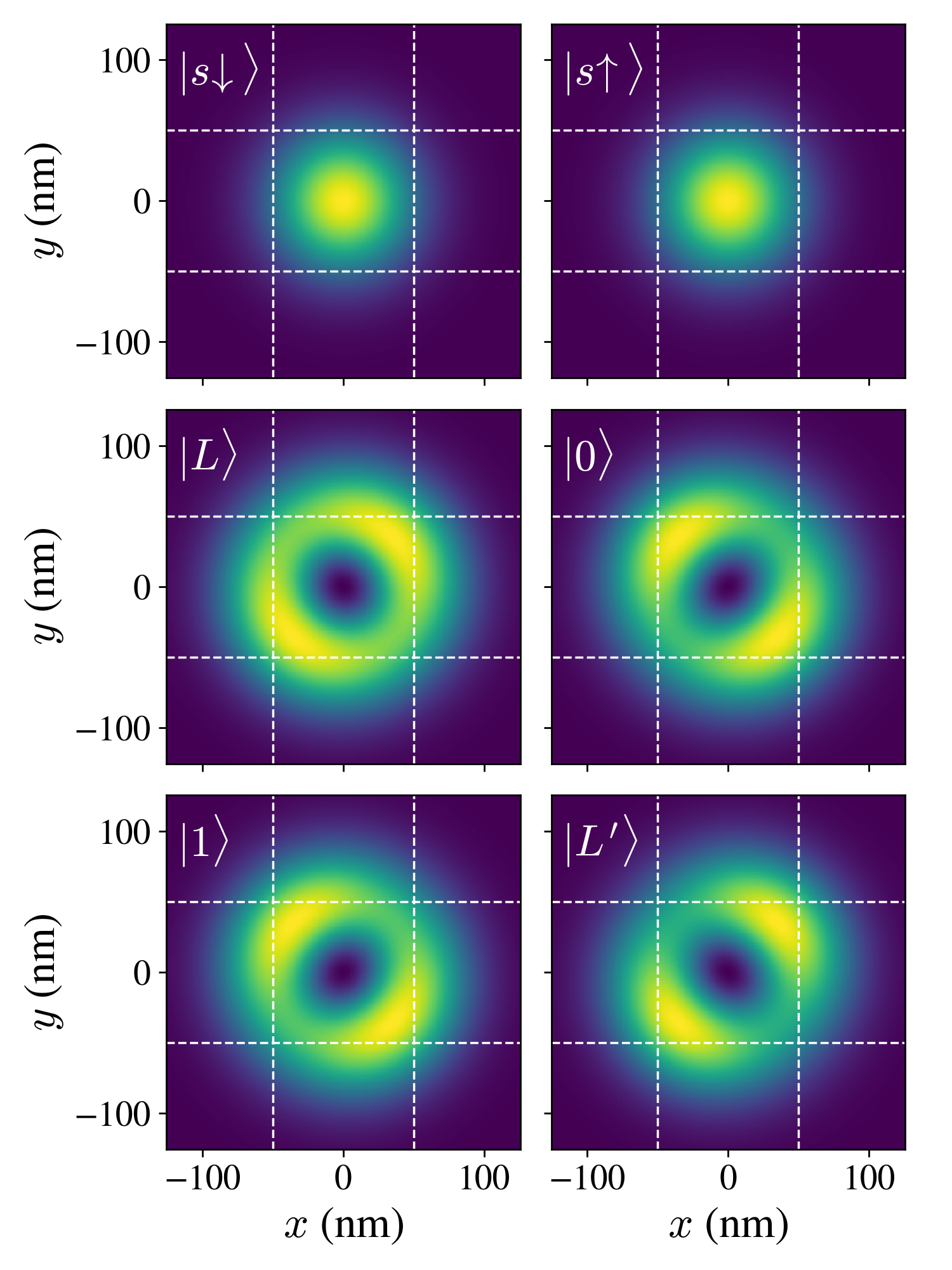}
    \caption{In-plane charge distributions of the $s$-shell doublet (top row) and of the four $p$-shell levels $\{\ket{L}, \ket{0}, \ket{1}, \ket{L'}\}$ (middle and bottom rows) at the operating point ($\varepsilon = 0$, $a_0 = 50$~nm, sweet-spot field orientation). Dashed lines mark $\pm a_0$.}
\label{fig:charge_dist}
\end{figure}

Figure~\ref{fig:charge_dist} shows the in-plane charge distributions obtained by summing over the spin component, i.e., $\int dz \sum_{J_z} |\Psi_n(x,y,z)|^2$, of the filled $s$-shell doublet and of the four $p$-shell levels $\{\ket{L}, \ket{0}, \ket{1}, \ket{L'}\}$ at the operating point $a_x^H=a_y^H=50~$nm, and $(B_\parallel,B_\perp)=(1~\mrm{T}, 7~\mrm{mT})$.
The $s$ states are Gaussian distributed, while all four $p$-shell levels form rings of nearly identical radius with a weak angular modulation along the diagonals set by the in-plane field direction of $45^\circ$.
In terms of the basis of Sec.~\ref{ssec:orbital_basis}, all four levels are composed almost entirely of the four HH $p$-orbital spinors $\{\ket{p_x,\pm3/2}, \ket{p_y,\pm3/2}\}$ [Fig.~\ref{fig:main}(b)], which carry $99.94\%$ of the norm in every case.
Their structure is most transparent in the circular basis $\ket{p_\pm} = (\ket{p_x} \pm i \ket{p_y})/\sqrt{2}$, which diagonalizes the in-plane orbital angular momentum $m_\ell = \pm 1$.
Retaining only the $p$-shell components, the numerical eigenvectors are
\begin{widetext}
\seq{
    \ket{L}  &\simeq 0.87\,e^{-i\pi/4}\ket{p_+\uparrow} + 0.06\,e^{i\pi/4}\ket{p_-\uparrow}
              - 0.50\,i\,\ket{p_+\downarrow}, \\[4pt]
    \ket{0}  &\simeq 0.39\,e^{-i\pi/4}\ket{p_+\uparrow} - 0.64\,e^{i\pi/4}\ket{p_-\uparrow}
              + 0.60\,i\,\ket{p_+\downarrow} - 0.27\,\ket{p_-\downarrow}, \\[4pt]
    \ket{1}  &\simeq 0.31\,e^{-i\pi/4}\ket{p_+\uparrow} + 0.59\,e^{i\pi/4}\ket{p_-\uparrow}
              + 0.61\,i\,\ket{p_+\downarrow} + 0.43\,\ket{p_-\downarrow}, \\[4pt]
    \ket{L'} &\simeq 0.03\,e^{-i\pi/4}\ket{p_+\uparrow} + 0.49\,e^{i\pi/4}\ket{p_-\uparrow}
              + 0.12\,i\,\ket{p_+\downarrow} - 0.86\,\ket{p_-\downarrow},
    \label{eq:eigenstate_decomp}
}
\end{widetext}
where $\uparrow\equiv+3/2$ and $\downarrow\equiv-3/2$ denote the HH spin projection.
All phases are exact multiples of $\pi/4$ at this operating point, a consequence of the in-plane field direction of $45^\circ$.
The omitted weight amounts to $0.06\%$ per level and lies predominantly in the LH band, with the largest single components appearing in the fourth LH out-of-plane subband at amplitudes below $1.2\times10^{-2}$.
There is no measurable $s$- or $d$-shell admixture, consistent with the deep central node visible in Fig.~\ref{fig:charge_dist}.
Equation~\eqref{eq:eigenstate_decomp} shows that $\ket{L}$ and $\ket{L'}$ are nearly pure circulating states, carrying $99.6\%$ and $98.5\%$ of their $p$-shell weight in $m_\ell = +1$ and $m_\ell = -1$ respectively, whereas $\ket{0}$ and $\ket{1}$ are close to equal superpositions of the two ($52\%/48\%$ and $47\%/53\%$).
This accounts for the ring-shaped densities of Fig.~\ref{fig:charge_dist}: a state of definite $m_\ell$ is azimuthally symmetric, and the residual angular modulation arises solely from the small counter-rotating admixture.
The outer levels are also predominantly spin-polarized, carrying $75\%$ ($\ket{L}$) or $76\%$ ($\ket{L'}$) of their $p$-orbital weight on a single HH spin projection, while the qubit states $\ket{0}$ and $\ket{1}$ are nearly equal spin mixtures ($57\%/43\%$ and $44\%/56\%$, respectively), reflecting the coherent spin-orbital mixing emphasized in Sec.~\ref{sec:quadrupole}.
The near-identical charge densities of the qubit states $\ket{0}$ and $\ket{1}$ underlie the weak coupling of the qubit to charge noise at the operating point and the weak state-dependence of the direct quadrupole-quadrupole coupling within the qubit subspace, as shown in Sec.~\ref{sec:quadrupole} and Sec.~\ref{sec:qubit-qubit}, respectively.

\section{$p$-orbital spectrum versus in-plane magnetic field angle}
\label{app-sec:in-plane-B-dependence}

\begin{figure*}[t]
    \centering
    \includegraphics[trim={0 0cm 0cm 0}, clip,width=0.85\textwidth]{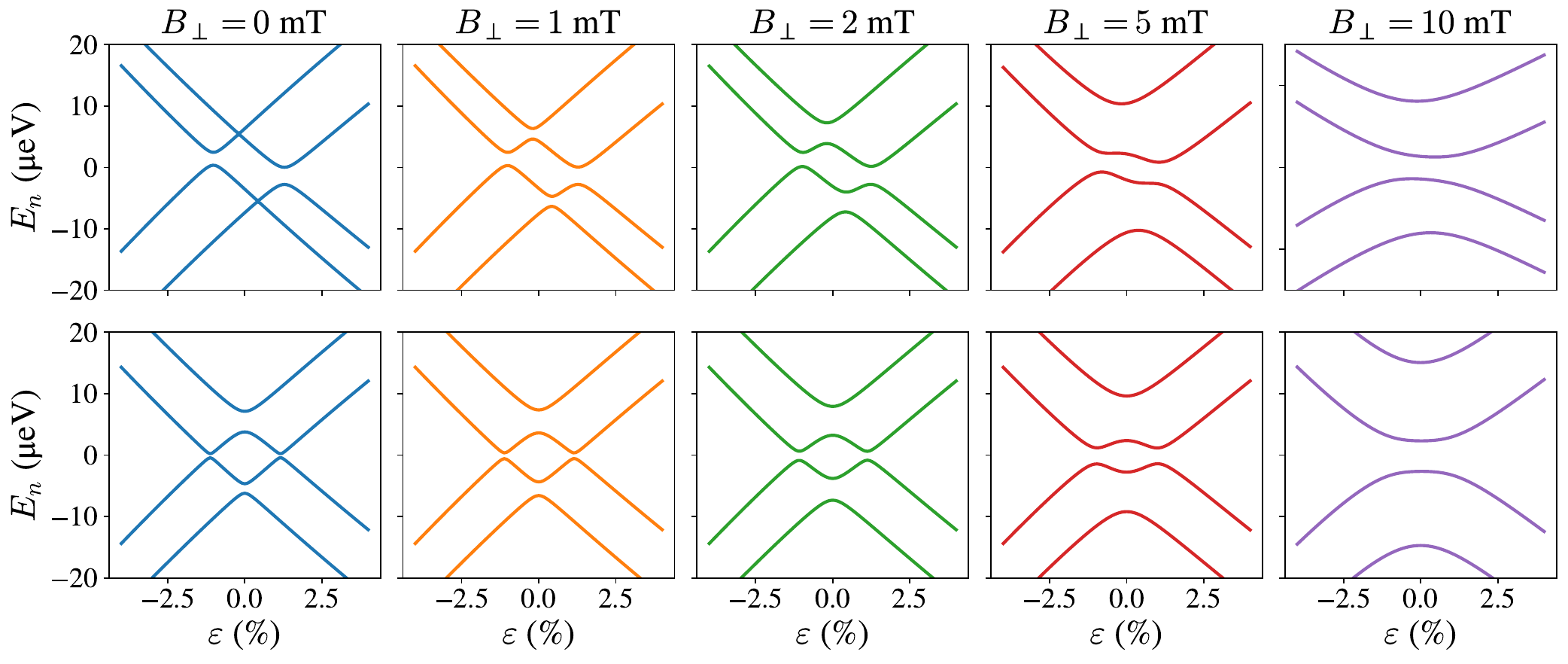}
\caption{$p$-shell energy spectra as a function of the dot anisotropy $\varepsilon$ for out-of-plane field components $B_\perp = 0$, 1, 2, 5, and 10~mT (columns). Top row: in-plane field along $x$, $\vec{B}=(B_\parallel,0,B_\perp)$. Bottom row: in-plane field at $45^\circ$, $\vec{B}=\l(B_\parallel/\sqrt{2},B_\parallel/\sqrt{2},B_\perp\r)$, the orientation used in the main text; $B_\parallel=1~$T in all panels. At $45^\circ$ the spectrum is symmetric under $\varepsilon \to -\varepsilon$ and flatter around the avoided crossing, motivating this choice of field orientation.}
\label{fig:Bangle}
\end{figure*}

Figure~\ref{fig:Bangle} shows the $p$-shell spectrum as a function of the anisotropy $\varepsilon$ for increasing out-of-plane field $B_\perp$, for two orientations of the in-plane field: along the $x$ axis (top row) and at $45^\circ$ between the $x$ and $y$ axes (bottom row), which corresponds to the orientation used throughout this work.
For the $45^\circ$ orientation, the spectrum is symmetric under $\varepsilon \to -\varepsilon$.
For a field along $x$ this symmetry is absent and the spectrum is visibly skewed with respect to $\varepsilon = 0$.
The symmetric orientation both centers the avoided crossing at the isotropic point and improves the extended flatness of the qubit branches around it, reducing the sensitivity of $\Delta_{01}$ to anisotropy fluctuations. 
This is the core property of the sweet-spot operation of Secs.~\ref{sec:quadrupole} and \ref{ssec:charge_noise}.
Note, in addition, that increasing $B_\perp$ opens the gap at the avoided crossing and ``smoothes-out'' some of the spectral features, consistent with the magnetic-field sweet-spot analysis of Fig.~\ref{fig:dephasing}.

\begin{figure}[t]
    \centering
    \includegraphics[trim={0 0cm 0cm 0}, clip,width=0.48\textwidth]{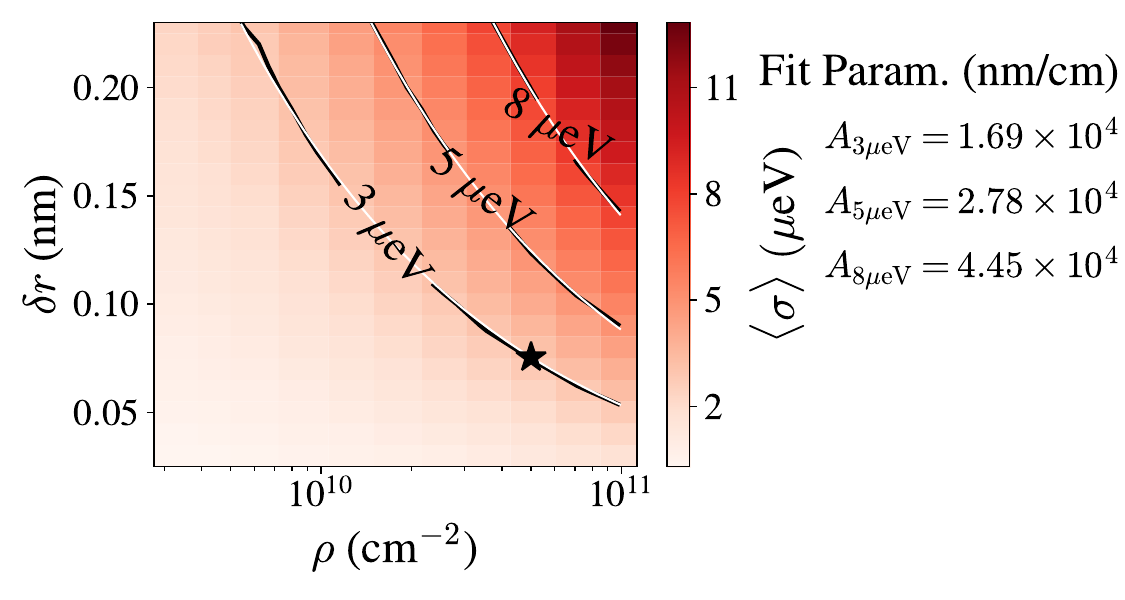}
\caption{TLF parameter calibration.
    Ensemble-averaged rms potential fluctuation $\langle\sigma\rangle$ as a function of the displacement magnitude $\delta r$ and charge density $\rho$, following Eq.~\eqref{eq:sigma_scaling}.
    The $\langle\sigma\rangle = 3\,\mu$eV contour identifies the physically relevant $(\rho,\delta r)$ pairs used in this work; the marked point denotes the working point $(\rho, \delta r) = (5\times10^{10}\,\mathrm{cm}^{-2},\, 0.075\,\mathrm{nm})$. The contour curves (black lines) are well described by the fitting ansatz $\delta r = A_\sigma / \sqrt{\rho}$ (white lines), where $A_\sigma=\langle\sigma\rangle/c_0$, compared with Eq.~\eqref{eq:sigma_scaling}.}
\label{fig:chemical_potential}
\end{figure}

\section{Calibration of the TLF ensemble}
\label{app:TLF}

In this section, we establish the calibration of the two free parameters of the TLF model: the areal density of active TLFs $\rho$ (in cm$^{-2}$), and the typical displacement magnitude $\delta r$ (in nm). 
Following Ref.~\cite{Kepa2023}, we determine the combinations $(\rho, \delta r)$ that reproduce a given fluctuation level of the dot's electrostatic potential.
For $1/f$ noise with PSD $S(f) = S_1/f$ and bandwidth $f_{\rm max}/f_{\rm min} = 10^{12}$ (corresponding to measurement times of $1\,\mu$s to $10^6\,\mathrm{s}$), the variance of the fluctuations of the ground state energy is
\seq{
    \sigma^2 &= 2 \int_{f_\mrm{min}}^{f_\mrm{max}} S(f) d f \\
    &= 2S_1 f_1 \ln\!\left(\frac{f_{\rm max}}{f_{\rm min}}\right) \\
    &\approx (3\,\mu\mathrm{eV})^2,
    \label{eq:sigma_1f}
}
where we have used $S_1 \approx 0.16\,\mu\mathrm{eV}^2/\mathrm{Hz}$, the noise amplitude at $f_1 = 1\,\mathrm{Hz}$ representative of state-of-the-art Ge/SiGe devices~\cite{Lodari2021,Hendrickx2024}. 
The analogous calibration for the Si/SiGe structure of Ref.~\cite{Kepa2023} uses $S_1 \approx 1\,\mu\mathrm{eV}^2/\mathrm{Hz}$, corresponding to $\sigma = 8\,\mu\mathrm{eV}$. 
This gives a reference noise scale of $\sigma = 3\,\mu\mathrm{eV}$, consistent with a sub-nanosecond $\tstar$ for charge qubits.

Here, $\sigma$ denotes the root-mean-squared fluctuation of the dot potential (chemical potential) induced by the TLF bath, and a Gaussian distribution of shifts is assumed. 
This quantity can be estimated directly in our particular system from the numerical diagonalization of the hole state, by analyzing the ground-state shift $\mel{0}{\delta V_n}{0}$ over many realizations of TLF spatial configurations (following the same steps outlined in Sec.~\ref{ssec:charge_noise}).
Thus, we define the fluctuation obtained from the TLF model by $\langle\sigma\rangle=\mrm{std}[\mel{0}{\delta V_n}{0}]$, shown in Fig.~\ref{fig:chemical_potential} for different values of $(\rho,\delta r)$.

Following Ref.~\cite{Kepa2023}, we use the following ansatz for the chemical potential fluctuations:
\seq{
    \langle\sigma\rangle    = c_0\,\sqrt{\rho} \, \delta r,
    \label{eq:sigma_scaling}
}
with fitted prefactor $c_0 \approx 1.795 \times 10^{-4}\,(\mathrm{cm/nm})\,\mu\mathrm{eV}$.
The contour lines in Fig.~\ref{fig:chemical_potential} identify the combination of $(\rho, \delta r)$ pairs consistent with this noise level at $\langle\sigma\rangle = 3,5,8\,\mu$eV. 
We select our working point on this contour by additionally requiring that the number of TLFs per frequency decade,
\seq{
    N_{\rm fpd}
    = \frac{N_{\rm TLF}}{\log_{10}(f_{\rm max}/f_{\rm min})},
    \label{eq:Nfpd}
}
is sufficiently large ($N_{\rm fpd} \gtrsim 2$--$3$) to avoid strong single-TLF Lorentzian features in the PSD, which are not representative of the smooth $1/f$ spectra observed in experiments~\cite{Kepa2023}. 
Choosing $\rho = 5\times10^{10}\,\mathrm{cm}^{-2}$ gives $N_{\rm TLF} = 47$ charges in the $300\,\mathrm{nm} \times 300~$nm simulation area ($N_{\rm fpd} \approx 3.9$) and a corresponding displacement magnitude $\delta r \approx 0.075~$nm.
These constitute the TLF parameters chosen in Sec.~\ref{ssec:charge_noise}.

\section{Dephasing time as function of anisotropy}
\label{app:T2-vs-anisotropy}

\begin{figure}[h!]
    \centering
    \includegraphics[trim={0 0cm 0cm 0}, clip,width=0.45\textwidth]{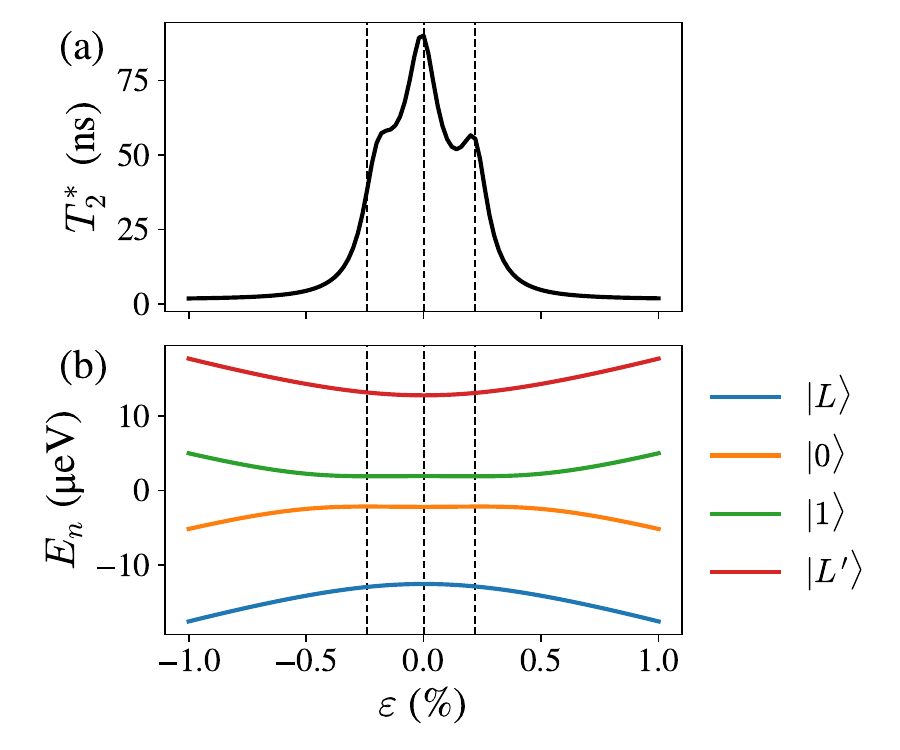}
    \caption{(a) Charge-noise-limited dephasing time $\tstar$ as a function of the dot anisotropy $\varepsilon$ at the sweet-spot field orientation, for $B_\parallel=1~$T, $B_\perp=7~$mT. (b) Corresponding $p$-shell spectrum. Dashed vertical lines mark the isotropic operating point $\varepsilon = 0$ and the two-qubit shuttle configuration $\varepsilon = \pm0.22\%$ (Sec.~\ref{sec:qubit-qubit}).}
\label{fig:T2eps}
\end{figure}

In this section, we study the dependence of $\tstar$ on dot ansiotropy $\varepsilon$.
Figure~\ref{fig:T2eps} shows the charge-noise-limited dephasing time $\tstar$ as a function of the dot anisotropy $\varepsilon$, computed with the TLF ensemble of Sec.~\ref{ssec:charge_noise}, together with the corresponding $p$-shell spectrum, for $(B_\parallel,B_\perp)=(1~\mrm{T},7~\mrm{mT})$.
$\tstar$ is maximal at the isotropic point and, remains comparable to its peak value within the flat region of the spectrum, directly quantifying the anisotropy tolerance of the sweet spot.
The dashed vertical lines mark: the isotropic operating point $\varepsilon = 0$, and the detuned configuration $\varepsilon = \pm0.22\%$ used during the two-qubit shuttle of Sec.~\ref{sec:qubit-qubit}, showing that the enhanced quadrupole-quadrupole coupling is obtained at only a modest cost in dephasing time.

\section{Phonon strain tensors and deformation-potential coupling}
\label{app:phonon}

An acoustic phonon of branch $\nu$ with wavevector $\vec{q} = q_\nu \hat{\mathbf{q}}(\hat{\Omega})$ and polarization unit vector $\hat{\mathbf{e}}^{\nu}(\hat{\Omega})$ produces a displacement field $\vec{u} \propto \hat{\mathbf{e}}^{\nu} e^{i\vec{q}\cdot\vec{r}}$, with associated strain $\epsilon_{ij} = (\partial_i u_j + \partial_j u_i)/2$~\cite{Li2020,Bulaev2005}. 
Factoring out the mode amplitude, which is absorbed in the prefactor of Eq.~\eqref{eq:Gamma_ph}, the dimensionless strain tensor entering the matrix element is
\seq{
    \l[ \epsilon_\nu(\hat{\Omega}) \r]_{ij} = \frac{1}{2} \l( \hat{q}_i\, \hat{e}^{\nu}_j + \hat{q}_j\, \hat{e}^{\nu}_i \r).
    \label{eq:phonon_strain}
}
In the isotropic approximation, the polarization vectors are $\hat{\mathbf{e}}^{l} = \hat{\mathbf{q}}$ for the longitudinal branch, and
\seq{
    \hat{\mathbf{e}}^{t_1} &= (\cos\theta\cos\varphi,\, \cos\theta\sin\varphi,\, -\sin\theta), \\
    \hat{\mathbf{e}}^{t_2} &= (-\sin\varphi,\, \cos\varphi,\, 0),
    \label{eq:phonon_pol}
}
for the two transverse branches, forming an orthonormal triad with $\hat{\mathbf{q}}(\hat{\Omega})$ of Sec.~\ref{ssec:relaxation}.

The deformation-potential coupling $H_\mrm{BP}[\epsilon]$ is the full Bir-Pikus Hamiltonian~\cite{BirPikus1974,Winkler2003}, with the same block structure as Eq.~\eqref{eq:HLK},
\seq{
    H_\mrm{BP}[\epsilon] =
    \begin{pmatrix}
        P_\epsilon + Q_\epsilon & 0 & S_\epsilon & R_\epsilon \\
        0 & P_\epsilon + Q_\epsilon & R_\epsilon^* & -S_\epsilon^* \\
        S_\epsilon^* & R_\epsilon & P_\epsilon - Q_\epsilon & 0 \\
        R_\epsilon^* & -S_\epsilon & 0 & P_\epsilon - Q_\epsilon
    \end{pmatrix},
    \label{eq:HBP_full}
}
with $P_\epsilon$ and $Q_\epsilon$ given in Eq.~\eqref{eq:PQeps}, and the shear blocks
\seq{
    R_\epsilon &= \frac{\sqrt{3}}{2}\, b_v \l( \epsilon_{xx} - \epsilon_{yy} \r) - i\, d_v\, \epsilon_{xy}, \\
    S_\epsilon &= -d_v \l( \epsilon_{xz} - i\, \epsilon_{yz} \r),
    \label{eq:RSeps}
}
where $d_v = -6.06$~eV is the shear deformation potential of Ge~\cite{Li2020}. 
For the static heterostructure strain of Sec.~\ref{ssec:hamiltonian} the shear components vanish and $H_\mrm{BP}$ reduces to the diagonal form of Eq.~\eqref{eq:HBP}. 
For phonon strain the shear blocks are in general nonzero and are retained in the evaluation of Eq.~\eqref{eq:Gamma_ph}.

\section{Two-hole multipole expansion and energy spectrum}
\label{app:2hole}

In this section, we specify the multipole expansion of Sec.~\ref{sec:qubit-qubit}, and provide the resulting $p$-level energy spectrum in the range of $R$ considered. 
With the A and B holes coordinates written as $\vec{r} = \vec{R}/2 + \vec{u}_A$ and $\vec{r}\,' = -\vec{R}/2 + \vec{u}_B$, measured from the two dot centers, the Coulomb kernel admits the Taylor expansion
\seq{
    \frac{1}{|\vec{r}-\vec{r}\,'|}
    = \sum_{k=0}^{\infty} \frac{1}{k!}
    \l[ \l(\vec{u}_B - \vec{u}_A\r)\cdot\nabla_{\vec{R}} \r]^k \frac{1}{R},
    \label{eq:multipole_series}
}
whose terms organize into products of dot-$A$ and dot-$B$ multipole operators: the term coupling the rank-$n$ moment of dot $A$ to the rank-$m$ moment of dot $B$ carries the interaction tensor $\partial_{i_1}\!\cdots\partial_{i_n}\partial_{j_1}\!\cdots\partial_{j_m} R^{-1} \propto R^{-(n+m+1)}$.
Through fourth order in the dot coordinates ($n+m \le 4$) the hierarchy is as follows.
The monopole-monopole term ($R^{-1}$) is the state-independent offset $e^2F_c/R$ of Sec.~\ref{sec:qubit-qubit}.
The dipole-monopole ($R^{-2}$), quadrupole-dipole and octupole-monopole ($R^{-4}$), and octupole-dipole ($R^{-5}$) terms all involve an odd-rank single-dot moment and are strongly suppressed by the near-parity of the $p$-shell eigenstates (Sec.~\ref{sec:quadrupole}); the dipole-dipole term ($R^{-3}$) carries this suppression twice.
The quadrupole-monopole ($R^{-3}$) and hexadecapole-monopole ($R^{-5}$) terms act on a single dot only and contribute calibratable single-qubit energy shifts.
The leading entangling contribution is therefore the quadrupole-quadrupole term of Eq.~\eqref{eq:UQQ}, which is the interaction retained in $H_{16}$.

\begin{figure}[t]
    \centering
    \includegraphics[trim={0 0cm 0cm 0}, clip,width=0.5\textwidth]{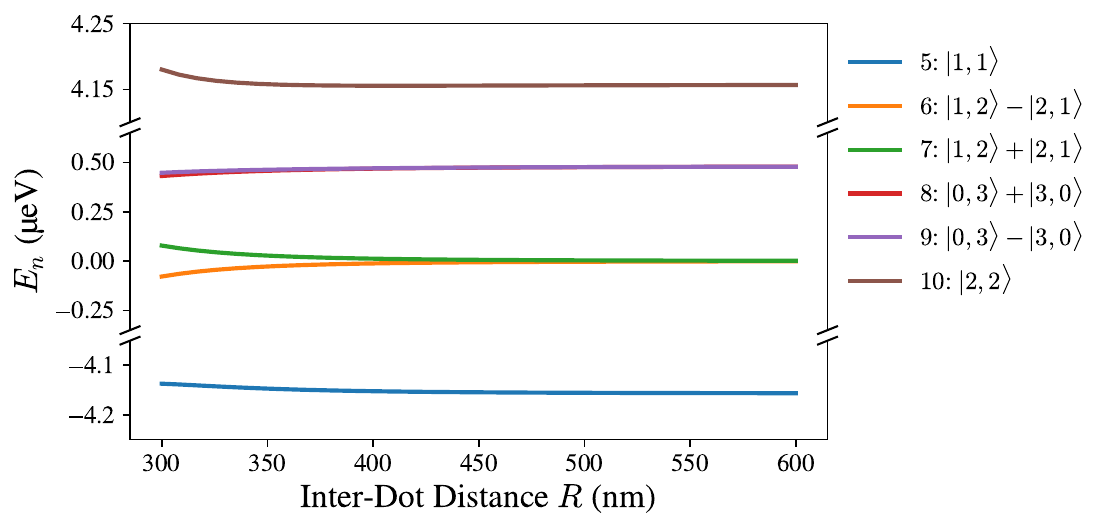}
\caption{Low-energy two-hole spectrum of $H_{16}(R)$ [Eq.~\eqref{eq:H2Q}] as a function of the inter-dot distance $R$. Eigenstates are labeled by their dominant product-basis component; solid coloring distinguishes the computational-subspace states from the leakage pair $(\ket{L,L'} \pm \ket{L',L})/\sqrt{2}$.}
\label{fig:2Q_spectrum}
\end{figure}

Figure~\ref{fig:2Q_spectrum} shows the spectrum near the computational states of the two-dot Hamiltonian $H_{16}[R]$ of Eq.~\eqref{eq:H2Q} as a function of the inter-dot separation $R$.
Shown as labels are the six eigenstates with dominant weight in the computational subspace and in the nearest leakage pair, labeled by their dominant product-basis component: the computational states $\ket{0,0}$, $(\ket{0,1} \pm \ket{1,0})/\sqrt{2}$, and $\ket{1,1}$, together with the leakage combinations $(\ket{L,L'} \pm \ket{L',L})/\sqrt{2}$. 
The remaining two-hole states are spectrally well separated from the computational subspace at all $R$ and are omitted from the figure. 
The leakage pair lies close in energy to the odd-parity qubit states and approaches them with decreasing $R$.
Nevertheless, the optimized shuttles of Sec.~\ref{sec:qubit-qubit} (and Appendix~\ref{app:ZZ}) leave these spectator states essentially unpopulated, as quantified by the unitarity of $U_4$ and the final gate fidelities.

\section{Direct optimization of a $ZZ$ target gate}
\label{app:ZZ}

\begin{figure}[t]
\centering
\includegraphics[width=\columnwidth]{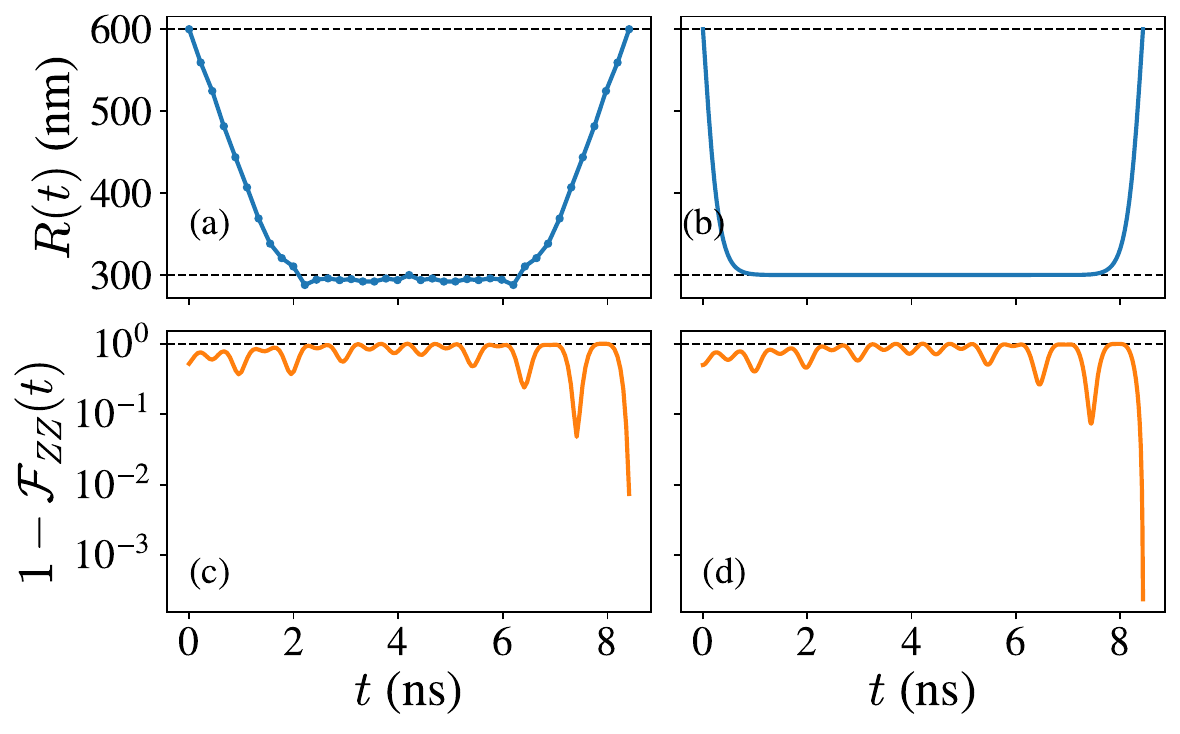}
\caption{(a),(b) Optimized inter-dot separation $R(t)$ for the direct $ZZ$-gate optimization between $R_{\max}=600~$nm and $R_{\min}=300~$nm: (a) GRAPE-style piecewise-constant-velocity optimization; (b) smooth flat-top ansatz of Eq.~\eqref{eq:shuttle_path}, with $t_f\approx8.4~$ns and $t_\mathrm{ramp}=0.3~$ns. (c),(d) Corresponding gate infidelity $1-\mathcal{F}_{ZZ}(t)$, with $\mathcal{F}_{ZZ}$ defined in Eq.~\eqref{eq:ZZ_fid}; note that $\mathcal{F}_{ZZ}(t_f)\approx 1$ for both.}
\label{fig:ZZ}
\end{figure}

As an alternative to the frame-independent optimization of Sec.~\ref{sec:qubit-qubit}, the shuttle can be optimized directly against the fixed target unitary. 
Here, we target the entangling gate $U_{ZZ} = \exp\l(i\tfrac{\pi}{4}\, ZZ\r)$. 
As a metric, we use the trace fidelity
\seq{
\label{eq:ZZ_fid}
\mathcal{F}_{ZZ}(t) = \l|\frac{1}{4} \Tr\l[U_{ZZ}^\dagger U_4(t)\r]\r|^2,
}
where $U_4(t)$ is the time-propagator projected onto the computational-subspace block of the full propagator $U_{16}(t)$, as in Sec.~\ref{sec:qubit-qubit}. 
The optimization minimizes the loss function $\mathcal{L}_{ZZ}=1-\mathcal{F}_{ZZ}(t_f)\exp\l[-(t_f/T_2^*)^2\r]$, whose quadratic dephasing penalty favors short gate times relative to the charge-noise-limited coherence time $T_2^*\approx100$~ns found in Sec.~\ref{ssec:charge_noise} for $\Bpar=1~$T.

As a benchmark, we performed a GRAPE-style optimization~\cite{Khaneja2005}, in which the shuttle velocity is taken piecewise-constant over 50 equal time segments, and the segment velocities are treated as independent control parameters.
In the GRAPE optimization, the shuttle velocity is initialized as constant, corresponding to a one-way traversal time $|\Delta R/\dot{R}(t)|=8~$ns, where $\Delta R=R_{\max}-R_{\min}$; this was found to be the shortest time where a solution is reachable via optimization.

In addition, we studied a shuttle path parametrized as a smooth double-sided flat-top profile,
\eq{
R(t) = R_{\max} - \Delta R
\begin{cases}
\tanh\!\l(\dfrac{t}{t_\mrm{ramp}}\r), & t \le t_f/2, \\[8pt]
\tanh\!\l(\dfrac{t_f-t}{t_\mrm{ramp}}\r), & t > t_f/2,
\end{cases}
\label{eq:shuttle_path}
}
with ramp time $t_\text{ramp}$ and total shuttle time $t_f$. 
This ansatz fixes $R(0)=R(t_f)=R_{\max}$ and reaches $R(t_f/2) \approx R_{\min}$ up to corrections that are exponentially small for $t_f \gg t_\text{ramp}$. 
The low-dimensional flat-top ansatz of Eq.~\eqref{eq:shuttle_path} was found to reproduce and exceed the fidelity of the much higher-dimensional GRAPE-style optimization, while yielding a smooth control waveform.

The two optimizations are compared side by side in Fig.~\ref{fig:ZZ} [left column, (a),(c): GRAPE; right column, (b),(d): flat-top ansatz].
Optimizing $(t_\text{ramp}, t_f)$ via Nelder-Mead yields $t_\text{ramp}=0.304$~ns and $t_f=8.437$~ns, achieving a coherent gate fidelity of $\mathcal{F}_{ZZ}(t_f)\approx 1$ [Figs.~\ref{fig:ZZ}(c) and \ref{fig:ZZ}(d)]. 
Including the dephasing penalty, the penalized fidelity $1-\mathcal{L}_{ZZ}$ reaches $99.36\%$, essentially saturating the limit set by the $T_2^*$ contribution to the cost function. 
The longer gate time compared to Sec.~\ref{sec:qubit-qubit} reflects the additional cost of matching a fixed frame: the optimizer must realize the target unitary exactly, including its local phases, rather than any member of its local-equivalence class.

\bibliographystyle{apsrev4-2}
\bibliography{refs}

\end{document}